\documentclass[useAMS,usenatbib,usegraphicx]{mn2e}
\usepackage{epstopdf}
\usepackage{ifpdf}
\usepackage{graphics,graphicx} 
\usepackage{epsfig} 

\newcommand{\gmip}{GMIP}         
\newcommand{\prep}{}   
\usepackage{grffile}
\usepackage{amssymb}
\title[Kinematics of Arp 270: gas flows, nuclear activity and two regimes of star formation]{Kinematics of Arp 270: gas flows, 
nuclear activity and two regimes of star formation}
\author[Zaragoza-Cardiel et al.]{J. Zaragoza-Cardiel$^{1}$ $^{2}$\thanks{E-mail:jzc@iac.es}, 
J. Font-Serra$^{1}$ $^{2}$, J. E. Beckman$^{1}$ $^{2}$ $^{3}$, J. Blasco-Herrera$^{4}$, \newauthor 
B. Garc\'ia-Lorenzo$^{1}$ $^{2}$, 
A. Camps$^{1}$ $^{2}$, O. Gonzalez-Martin$^{1}$ $^{2}$, C. Ramos Almeida$^{1}$ $^{2}$, \newauthor 
N. Loiseau$^{5}$ and L. Guti\'errez$^{6}$\\
$^{1}$Instituto de Astrof\'isica de Canarias, C/ V\'ia L\'actea s/n, 38205 La Laguna, Tenerife, Spain\\
$^{2}$Department of Astrophysics, University of La Laguna, E-38200 La Laguna, Tenerife, Spain\\
$^{3}$CSIC, 28006 Madrid, Spain\\
$^{4}$Instituto de Astrof\'isica de Andaluc\'ia (CSIC) Apartado 3004, 18080 Granada, Spain\\
$^{5}$ XMM-Newton Science Operations Centre, ESAC/INSA, Villafranca del Castillo, Spain\\
$^{6}$Universidad Nacional Aut\'onoma de M\'exico, Instituto de Astronom\'ia, Ensenada, B. C., Mexico}

\begin{document}

\label{firstpage}

\maketitle

\begin{abstract}
We have observed the Arp 270 system (NGC 3395 \& NGC 3396) in H$\alpha$ emission using the GH$\alpha$FaS Fabry-Perot 
spectrometer on the 4.2m William Herschel Telescope (La Palma). In NGC 3396, which is edge-on to us, we detect gas inflow 
towards the centre, and also axially confined opposed outflows, characteristic of galactic superwinds, and we 
go on to examine the possibility that there is a shrouded AGN in the nucleus. 
The combination of surface brightness, velocity and velocity dispersion information enabled us to measure the radii, 
FWHM, and the masses of 108 HII regions in both galaxies. We find two distinct modes of physical behaviour, 
for high and lower luminosity regions. We note that the most luminous regions 
show especially high values for their velocity dispersions and hypothesize that these occur because 
the higher luminosity regions form from higher mass, gravitationally bound clouds while those at lower luminosity HII regions 
form within molecular clouds of lower mass, which are pressure confined.

\end{abstract}

\section{Introduction}

\begin{figure*}

\centering
\includegraphics[width=0.85\linewidth]{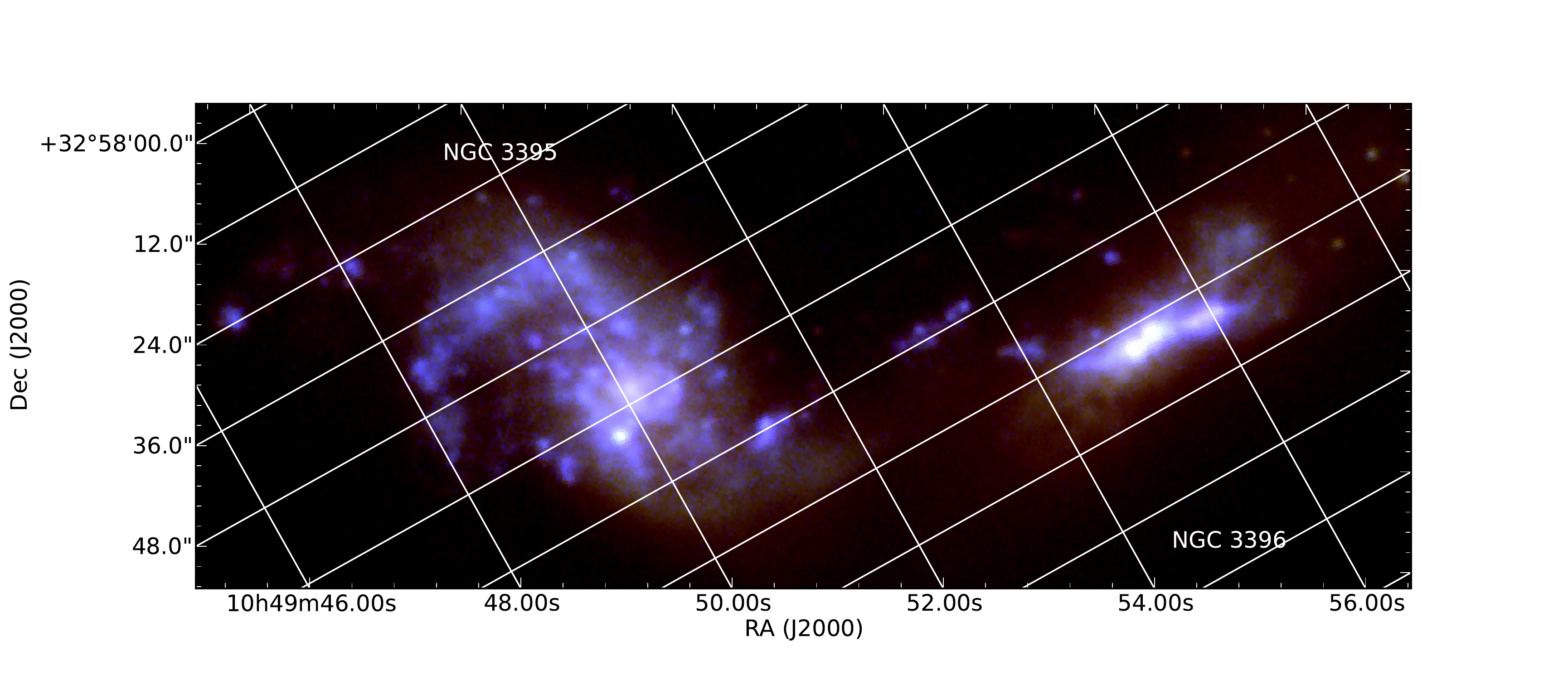}

\caption{Composite colour map of Arp 270 from SDSS images. In this figure, and in all subsequent figures with 
images of Arp 270, the coordinates axes are those displayed explicitly in this figure.}
\label{fig:figureSDSS}
\end{figure*}

The detailed effects of galaxy interactions on 
nuclear activity \citep{Canalizo07,Bennert08,Georgakakis09,2011ApJ...726...57C,Ramos11,Ramos12,Bessiere12} 
and also on star formation 
\citep{Somerville08,2011EAS....51..107B,Tadhunter11}
demand considerable further study. The scaling relations showing 
the dependence of emission line widths in molecular clouds appear to be determined by supersonic 
turbulence \citep{1999osps.conf....3B}. Turbulence gives rise to density perturbations, 
locally triggering gravitational collapse, and leading to the formation of stars \citep{2004RvMP...76..125M, 
2011EAS....51..133K}. But there is also evidence that magnetic fields play a role in controlling the collapse 
\citep{0004-637X-750-1-13}. The relative importance of these effects, in general, is not well understood, 
so that a variety of new observational results is needed to help clarify them. There is also evidence for 
the existence of triggered star formation. As pointed out by \citet{2011EAS....51..107B}, 
the density is a parameter which is 
important in determining how the stars form, so that if we have sets of placental clouds with different 
density regimes we might well expect distinct star formation regimes within them.
We have initiated a program of kinematic observations of interacting galaxy pairs analyzing the H$\alpha$ emission line, 
using an instrument which gives unequalled angular and velocity resolution per unit observation time. These observations, 
with the GH$\alpha$FaS Fabry-Perot 2D spectrometer on the 4.2m William Herschel telescope, allow us, \emph{inter alia}, 
to explore a part of HII region parameter space normally not attainable in all but the most massive 
single galaxies: that of the most luminous regions (in galaxies of lower mass such a 
study is limited by low number statistics). They also allow us more generally to gain insights into 
how the dynamical effects demonstrated by the kinematic disturbances affect properties such as the presence of an AGN, 
and the overall SFR. A set of results for the weakly interacting pair in the Arp 271 system have 
recently been presented \citep{2011ApJ...740L...1F}, in which gas transfer detected between the two galaxies 
could be quantitatively related to an enhanced SFR and a 
nuclear wind in one of them. In the present article we present our 
observations of the Arp 270 system, which contains an apparently almost face-on galaxy, NGC 3395, and an almost edge-on galaxy,
NGC 3396 (Figure \ref{fig:figureSDSS}), and which has reached a phase of closer interaction than the galaxies in Arp 271.

The distance to Arp 270 is $21.8 \, \mathrm{Mpc}$ \citep{1992ApJS...80..479T}, yielding a scale of 
$0.1\, \mathrm{kpc}\, \mathrm{arcsec}^{-1}$. The systemic
velocities of NGC 3395/6 are $1605\, \mathrm{km}\, \mathrm{s}^{-1}$ and $1675\, \mathrm{km}\, \mathrm{s}^{-1}$ 
while the inclinations are $50^{\circ}$ and $70^{\circ}$ respectively 
\citep{2002A&A...387..821G}. 

Arp 270 is an approximately equal-mass merger (mass ratio 1.5:1) thought to be at its 2nd approach, e.g.
\cite{1999MNRAS.308..364C}, its star formation rate remainS moderate
($L_{FIR} \lesssim 10^{44}\, \mathrm{erg/s}$ \citep{2003AJ....126.1607S}, thus $SFR \lesssim5\, \mathrm{M_{\odot}/yr}$)
but is probably due to increase at the
forthcoming pericenter since it is in a pre-starburst rather than post-starburst condition.


In section $\S2$ we explain the observations and the calibration method used.  In section $\S3$ 
we present an analysis of the global kinematics, focusing on the most interesting case of 
NGC 3396 plus the possibility that there is an AGN in it, 
which may be fed by the interaction. In section $\S4$ we show the physical properties inferred 
for the HII regions in the galaxies, with evidence of two physically distinct populations. 
We also offer suggestions for the origin of the two regimes. In section $\S5$ we give a 
discussion and present conclusion.

\section{Observations \& H$\alpha$ Calibration}

\begin{figure*}

\centering
\includegraphics[width=0.85\linewidth]{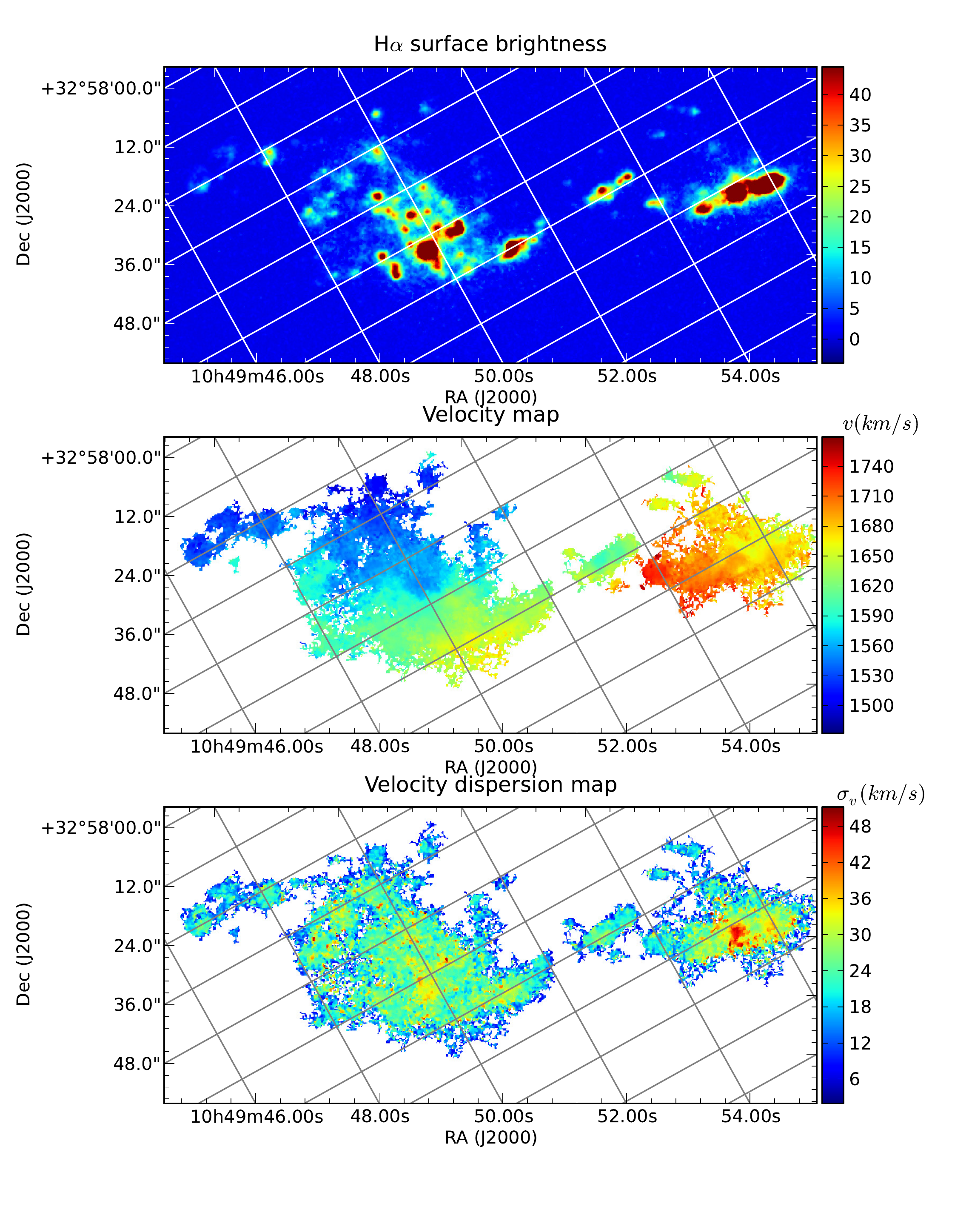} 

\caption{(From top to bottom: H$\alpha$ surface brightness; 
first moment map, giving the velocity fields of the two galaxies forming Arp 270; 
second moment map showing the velocity dispersion.}
\label{fig:figure_kin}
\end{figure*}

The observations were taken on the night of December $23^{rd}$, 2010, with GH$\alpha$FaS (Galaxy H$\alpha$ 
Fabry-Perot Spectrometer, \cite{2008PASP..120..665H}, \cite{2008ApJ...675L..17F}) , at the Nasmyth Focus of 
the 4.2m William Herschel Telescope, ORM (La Palma). The spectrometer 
has a circular FOV with a diameter of 3.5 arcmin. In H$\alpha$, the etalon used covers a Free Spectral 
Range (FSR) of 8$\, \mathrm{\AA}$ which corresponds 
to 390 km/s with a spectral resolution $\delta \lambda$ of 0.2$\, \mathrm{\AA}$ which corresponds to a velocity resolution 
$\delta v$ of some 8 km/s. The pixel size of the detector subtends 0.2 arcsec on the sky, and the angular resolution 
was seeing limited at around 1 arcsec or 0.1 kpc on the galaxy assuming a distance of 21.8 Mpc. 
As the instrument operates at the Nasmyth focus, and GH$\alpha$FaS has not an optical derotator, 
the observations are affected by rotation of the field. However, 
GH$\alpha$FaS takes individual spectral scans covering the complete spectral range over the full field in only  8  min 
which allows us to use a digital derotation technique to bring all the scans into a uniform 
geometrical register. The technique has been described in \cite{2010MNRAS.407.2519B}. After applying this correction, 
calibration in velocity, and phase adjustment we have a data cube to which we have applied the 
procedures described in \cite{2006MNRAS.368.1016D} to derive maps of Arp 270 in H$\alpha$ surface brightness, velocity, 
and velocity dispersion, which are shown in Figure \ref{fig:figure_kin}. 

In order to calibrate the H$\alpha$ in flux we used the measured flux for the system obtained by \cite{2009ApJ...703.1672K} 
and compared it with the integrated photon count per second from GH$\alpha$FaS, $F_G$, once corrected for the specific 
filter transmittance, that is: 


\begin{equation}
F_{H\alpha}\, \mathrm{(W/cm^2)}=7.7\cdot 10^{-21}\thinspace F_{G}\, \mathrm{(counts/s)}
\label{eq:eq_cal} 
\end{equation}

Although 
using the galaxies as a single object does entail some uncertainty, the results are consistent, within 10\%, with those 
of \cite{2013blasco}, and with calibrations performed on individual regions in M61 using 
the data from \cite{2004A&A...426.1135K} and data observed with GH$\alpha$FaS.

\begin{equation}
F_{H\alpha}\, \mathrm{(W/cm^2)}=8.6 \cdot 10^{-21}\thinspace F_{G}  - 2.17\cdot 10^{-21}\, \mathrm{(counts/s)}
\label{eq:eq_calm61} 
\end{equation}

The previous flux calibrations ignore the effect of dust attenuation. 
Unfortunately, we are not be able to correct of dust attenuation for each region of the galaxies, but we can 
estimate the dust attenuation in each galaxy from their 1.4GHz luminosities 
given in \cite{2002AJ....124..675C}. Then, using the calibration of \cite{2009ApJ...703.1672K} (equation 12) to correct 
for dust attenuation using 1.4GHz luminosity, we have found that the global 
SFR for NGC3396 needs to be corrected by a factor of $2.3$ and NGC3395 needs to be 
corrected by a factor of $3.5$. However, as we study specific regions of the 
galaxies, we do not have the data to correct each of them for attenuation, so the values we 
give are lower limits to the SFR. 
It is significant that 
NGC3395 is more affected by dust attenuation than NGC3996 as we will comment below.

\section{Global kinematics}

The velocity map in Figure \ref{fig:figure_kin} shows a distorted rotation field in NGC 3395, and an apparently 
similarly distorted field in NGC 3396. However the velocity range in the latter case, $\Delta v\sim80\, \mathrm{km/s}$, is much too 
small for an apparently edge-on galaxy with the mass which we have found in the literature \citep{1999MNRAS.308..364C} of 
$\sim 2\cdot10^{10}\, \mathrm{M_{\odot}}$ 
and calls for more detailed analysis which we present here. The velocity dispersion map
in \ref{fig:figure_kin} shows quite high velocity dispersions (up to $\sim 50km/s$) and a correlation
between the peaks of the velocity dispersions and the star-forming knots. We will present these aspects in more
detail in section $\S4$. These maps and the H$\alpha$ datacube are available through CDS. 

Although the galaxies are clearly affected by the interaction, we have estimated the kinematical parameters using 
the velocity maps. We did this 
by using ADHOC\footnote{\emph{http://www.oamp.fr/adhoc/}} software, written by J. Boulesteix, 
(Observatoire de Marseille), to optimize the circular velocity field. We show the results in Table \ref{table:kin}, where we 
have included the calibration of H$\alpha$ in flux and the position angle (PA) is defined 
as ``the counter-clockwise angle between north 
and a line which bisects the tracer's velocity field, measured on the receding side'' \citep{2011MNRAS.417..882D}.

\begin{table*}
\centering
\begin{tabular}{|c|c|c|c|c|c|c|}
\hline
Object & $log(L_{H\alpha})\, \mathrm{(erg/s)}$ & ${Centre}_{kin} $  (RA;Dec)& $PA$ ($^{\circ}$) & $i$ ($^{\circ}$) & $v_{sys} \, \mathrm{(km/s)}$ \\
\hline
\hline
NGC 3395& $41.34$ & $ 10\, \mathrm{h}\thinspace49\, \mathrm{m}\thinspace50.31\, \mathrm{s}\thinspace\pm\thinspace0.07\, \mathrm{s};\thinspace 32^{\circ}59'00''\thinspace\pm\thinspace1'' $    &   $1\thinspace\pm\thinspace6$ & $59\thinspace\pm\thinspace3$ & $1600\thinspace\pm\thinspace 20$\\

\hline
NGC 3396& $40.99$&$ 10\, \mathrm{h}\thinspace49\, \mathrm{m}\thinspace55.30\, \mathrm{s}\thinspace\pm\thinspace0.07\, \mathrm{s};\thinspace 32^{\circ}59'26''\thinspace\pm\thinspace1'' $    &   $286\thinspace\pm\thinspace6$ & $67\thinspace\pm\thinspace4$ & $1690\thinspace\pm\thinspace 40$\\

\hline
\end{tabular}
\caption{Kinematic parameters.}
\label{table:kin}
\end{table*}     

\subsection{Gas inflow towards the nucleus of NGC 3396}
\begin{figure}
 \centering
\begin{tabular}{cc}

\includegraphics[width=0.45\linewidth]{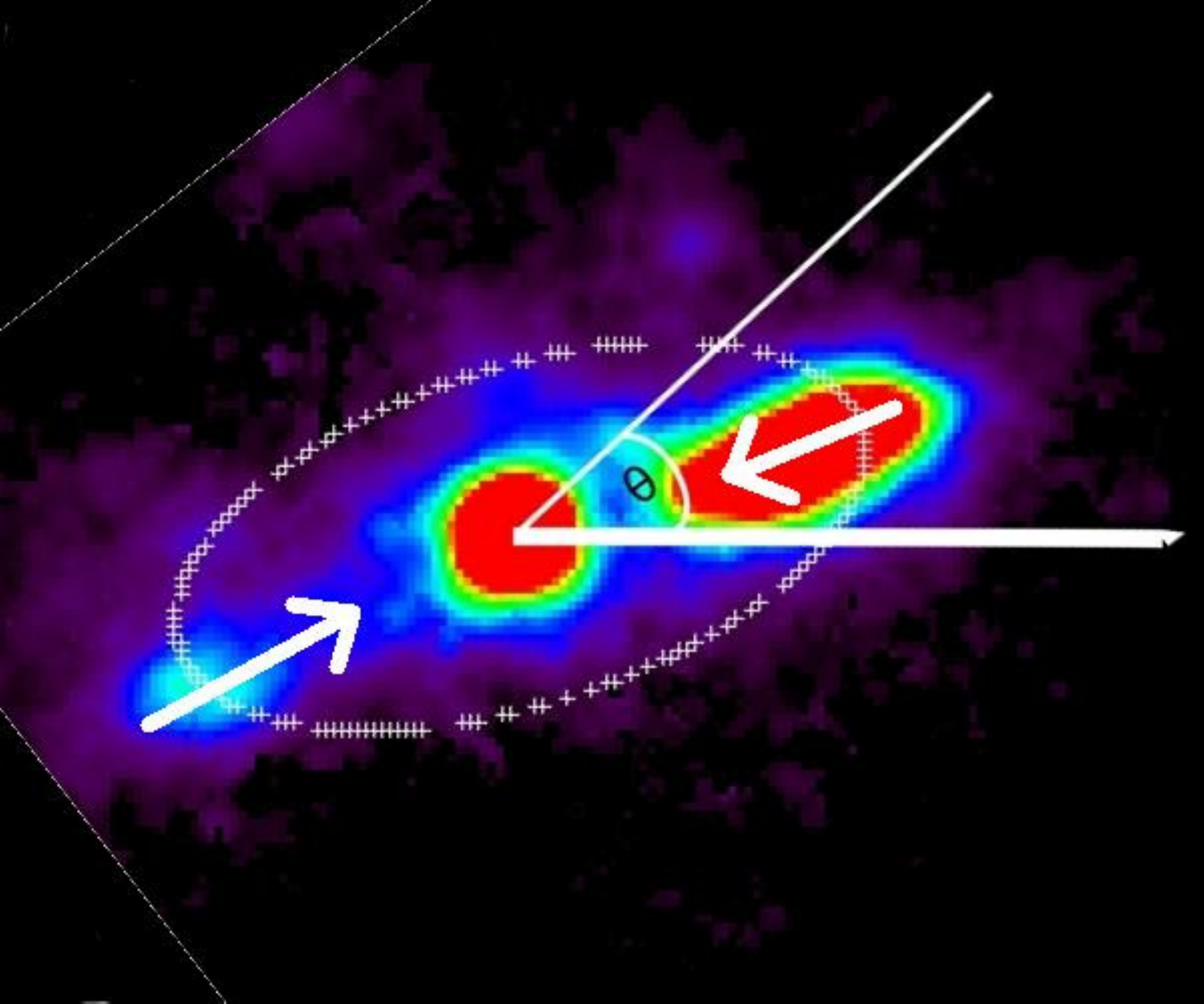} &

\includegraphics[width=0.45\linewidth]{./figuras/elipse_density_velocity_low.pdf}\\
\end{tabular}

\caption{H$\alpha$ surface brightness (top-right) and radial velocity (bottom-right) plotted in an ellipse (left) 
around the nucleus}
\label{fig:figure2}

\end{figure}

It is well known that interactions between galaxies can induce inflows of gas towards their 
centres (see e.g. \cite{1996ApJ...464..641M}). We can see from the H$\alpha$ surface brightness map in Figure \ref{fig:figure_kin} 
that there is a powerful source of emission in the central zone of NGC 3396. This leads directly to a possible 
explanation for the low range of ``rotational`` velocities detected in this galaxy: they are not in fact rotational, 
but projected inflow velocities along what appears to be a bar. Since NGC 3396 was classified as a barred galaxy by 
\cite{1991rc3..book.....D} so we do not need to claim the presence of a bar but the non-rotational velocities. 
In order to test this scenario we have combined the 
H$\alpha$ surface brightness and velocity information by plotting their values around elliptical loci, which are 
the projections on the sky of circular orbits. An example is shown in Figure \ref{fig:figure2}, where the velocity gradient 
is enhanced as we cross the bar; on one side the velocity component towards us decreases, and that 
towards the nucleus increases, and on the other side the effect is reversed. 

This behaviour is detected along the whole length of the bar. The velocity profiles are strongly reminiscent 
of the simulated profiles in \cite{1992MNRAS.259..345A}, in which gas is shocked as it collides with the bar, 
forming dust lanes, with their characteristic morphology.

On the assumption that we can derive the underlying kinematic parameters using the distorted velocity map of NGC 3396, 
we then went on to obtain the non-circular component, which in this case is the inflow velocity.  
We have subtracted off the rotation field obtained while estimating global kinematics parameters (Table \ref{table:kin}) 
from the 
observed field, to yield a map of the residual, non-circular velocities. 

\begin{figure*}
 
\centering


\includegraphics[width=0.75\linewidth]{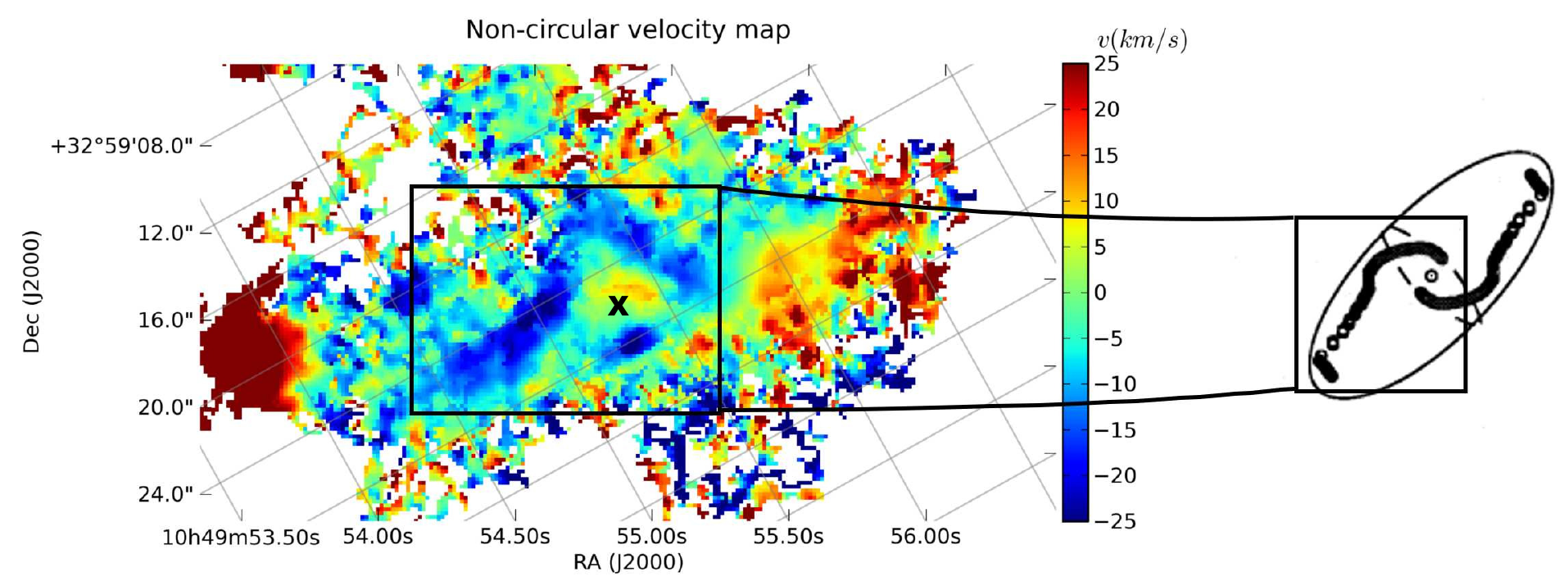}

\caption{Left, non-circular velocity map of NGC3396, 
the ''X'' mark represents the 
kinematic nucleus. Right, dust lane model from \citet{1992MNRAS.259..345A}}

\label{fig:figure0}
\end{figure*}

The resulting map is shown Figure \ref{fig:figure0}, 
where the negative components of the non-circular motion (blue regions) to the left of the kinematic nucleus,
 and the positive components (yellow regions) to the right of the nucleus have the shape of typical 
curved dust lanes observed in barred galaxies \citep{1984PhR...114..319A}.
We have been able to make a zero order estimate of $\phi$, the ionized gas mass inflow rate, using the 
approximation $\phi = 2\cdot \rho\cdot \sigma  \cdot v$. In this expression $\rho$ is the ionized gas density, which we 
have estimated using the mean luminosity-weighted electron density, hereinafter for 
simplicity called electron density, $n_e=<n_e>$; $\sigma$ is the cross 
section of the inflow along the bar, and $v$ is the velocity directed towards the nucleus. We have incorporated a 
factor 2 to take into account the flow along both sides of the bar. Because of the geometry of NGC 3396, 
to one side of the bar what we first see in the line of sight is the dust (blue regions in 
Figure \ref{fig:figure0}). Thus, we are able to observe the geometry on this side of the bar and this is the 
side we have used to estimate the velocity of the inflow and the diameter of an idealized cylinder where the inflow occurs. 
However, as the dust is absorbing the H$\alpha$ emission, we have to estimate the electron density from the other 
side of the bar, where the star formation is seen before the dust in the line of sight.

Following \cite{2005A&A...431..235R} we can measure the electron density, 
$n_e=\rho$, using the H$\alpha$ luminosity and an estimate of the dimensions of the emitting region. 
If we consider spherical HII regions composed only by hydrogen with uniform density \citep{1978ppim.book.....S}:


\begin{equation}
\frac{L_{H\alpha}}{\pi R^2}=h\nu_{H\alpha}\alpha_{H\alpha}^{eff}(H_0,T)2.46\cdot 10^{17}\cdot n_e^2R
\label{eq:density}
\end{equation}

where $\alpha_{H\alpha}^{eff}(H_0,T)$ is the effective recombination coefficient of the $H\alpha$ emission line, 
$h\nu_{H\alpha}$ is the energy of an $H\alpha$ photon and R is the radius in cm of the HII region. 

We estimate $n_e$ from the value of the electron density found in the approaching part of the bar 
in section \S4  
yielding $\rho=n_e=4.2\, \mathrm{e^-/cm^3}$.

From the non-circular velocity map, 
we estimate the cross section diameter using the limits to the zone where the non-circular velocity is 
less than -10 km/s (since the uncertainty in our velocity measurements is $~8\, \mathrm{km/s}$), which yields 
an approximate value of $D=(300\pm100)\, \mathrm{p}c$.

If $(r, \Phi)$ and $(R,\theta)$ are the polar coordinates in the sky and galaxy planes respectively, 
the $v_{los}$ can be written \citep{1978ARA&A..16..103V} 

\begin{equation}
 v_{los}=v_{\theta}\thinspace sin\thinspace i \thinspace cos \thinspace \theta + v_{R}\thinspace sin\thinspace i \thinspace cos \thinspace \theta
\label{equation:project}
\end{equation}

where $v_{\theta}$ and $v_R$ are the circular and the radial velocity (velocity towards the nucleus, $v_{in}$, 
in our case) respectively. The relation between the galaxy and sky coordinates is:

\begin{equation}
 tg\thinspace \theta=\frac{tg\thinspace(\Phi-\Phi_0)}{cos\thinspace i}
\label{equation:project_angle}
\end{equation}

\begin{equation}
R=\frac{r\thinspace cos \thinspace (\Phi-\Phi_0)}{cos\thinspace \theta} 
\label{equation:project_radi}
\end{equation}

As we already have removed the circular velocity, $v_{\theta}=0$ for the residual map (Figure \ref{fig:figure0}). 
Then, $v_R$ projected in the line of sight, $v_{los}$, is  
the one we can measure in the residual map. We estimate $v_{los}$ and $\Phi-\Phi_0$ 
using the side of the bar where the line of sight first mets the dust, 
 as in our estimate of the diameter.

The mean value of the $v_{los}$ for the 
points where $v_{los}>10\, \mathrm{km/s}$ is $<v_{los}>=(17.0\pm4.0) \, \mathrm{km/s}$. The result for the angle between the bar and the kinematic 
position angle is $<\Phi-\Phi_0>=(27\pm8)^{\circ}$ and the inclination obtained in the determination of the rotation curve, 
$i=(67\pm4)^{\circ}$, implying $\theta=(47\pm15)^{\circ}$. Combining the previous results with 
equations \ref{equation:project}, \ref{equation:project_angle} 
and \ref{equation:project_radi} we obtained an inflow velocity of $v_{in}=(27\pm15)\, \mathrm{km/s}$.
 



The ionized gas inflow rate result is a value for $\phi$ of $(0.4\pm0.2)\, \mathrm{M_{\odot}/yr}$, along the bar. 
From the H$\alpha$ emission in the circumnuclear zone, using the calibration of \cite{2009ApJ...703.1672K} we find a 
star formation rate of $0.15 \, \mathrm{M_{\odot}/yr}$. These two values are mutually consistent: there appears to be 
enough inflow to fuel the circumnuclear star formation.

\begin{figure}

\centering
\includegraphics[width=0.9\linewidth]{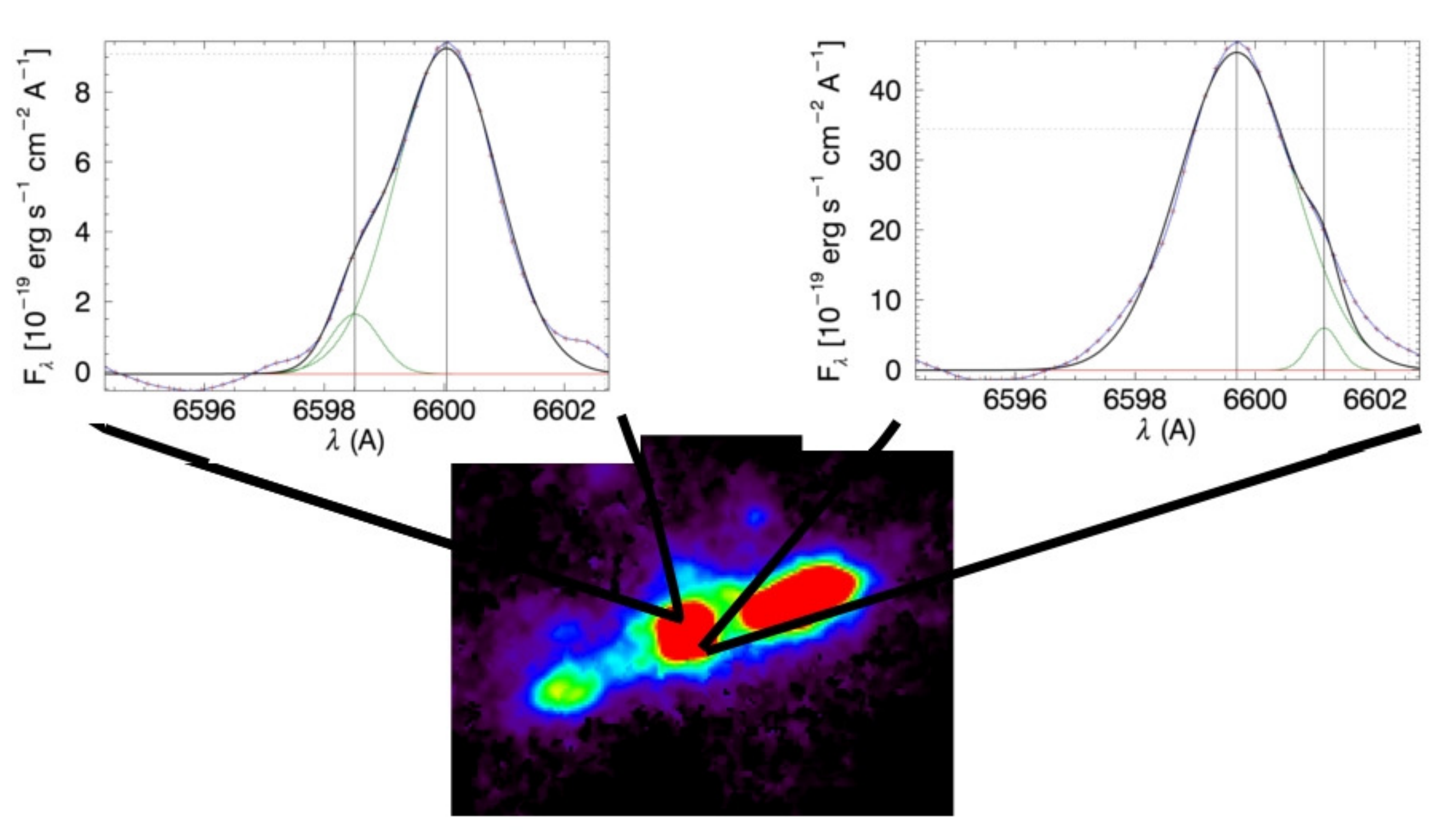}

\caption{H$\alpha$ profile in opposite sides of the nucleus of NGC 3396}
\label{fig:figure3}

\end{figure}

\subsection{Outflow from the nucleus of NGC 3396}

H$\alpha$ line profiles obtained in our GH$\alpha$FaS data present a variety of shapes showing multiple peaked 
profiles close to the nucleus (see Figure \ref{fig:figure3}).
The position of the secondary peak with 
respect to the main peak shifts as we change position around the nucleus, and the form of these 
shifts can be explained naturally in terms of a biconical outflow in a direction perpendicular to the plane of 
the inflow. This is similar to previous studies by \cite{1998A&A...330..841C} who found biconical outflow 
from a central bar-like structure in the interacting galaxy UGC 3995, as well as evidence for inflow along the bar, 
and by \cite{2001AJ....121..198V}, who found a biconical outflow from the AGN of NGC 2992. 

Due 
to the small separation between 
components a simple multiple gaussian fit does not work without a proper values for initial parameters 
(the solution is strongly dependent on the initial values). 

\begin{figure}
\centering

\includegraphics[width=0.9\linewidth]{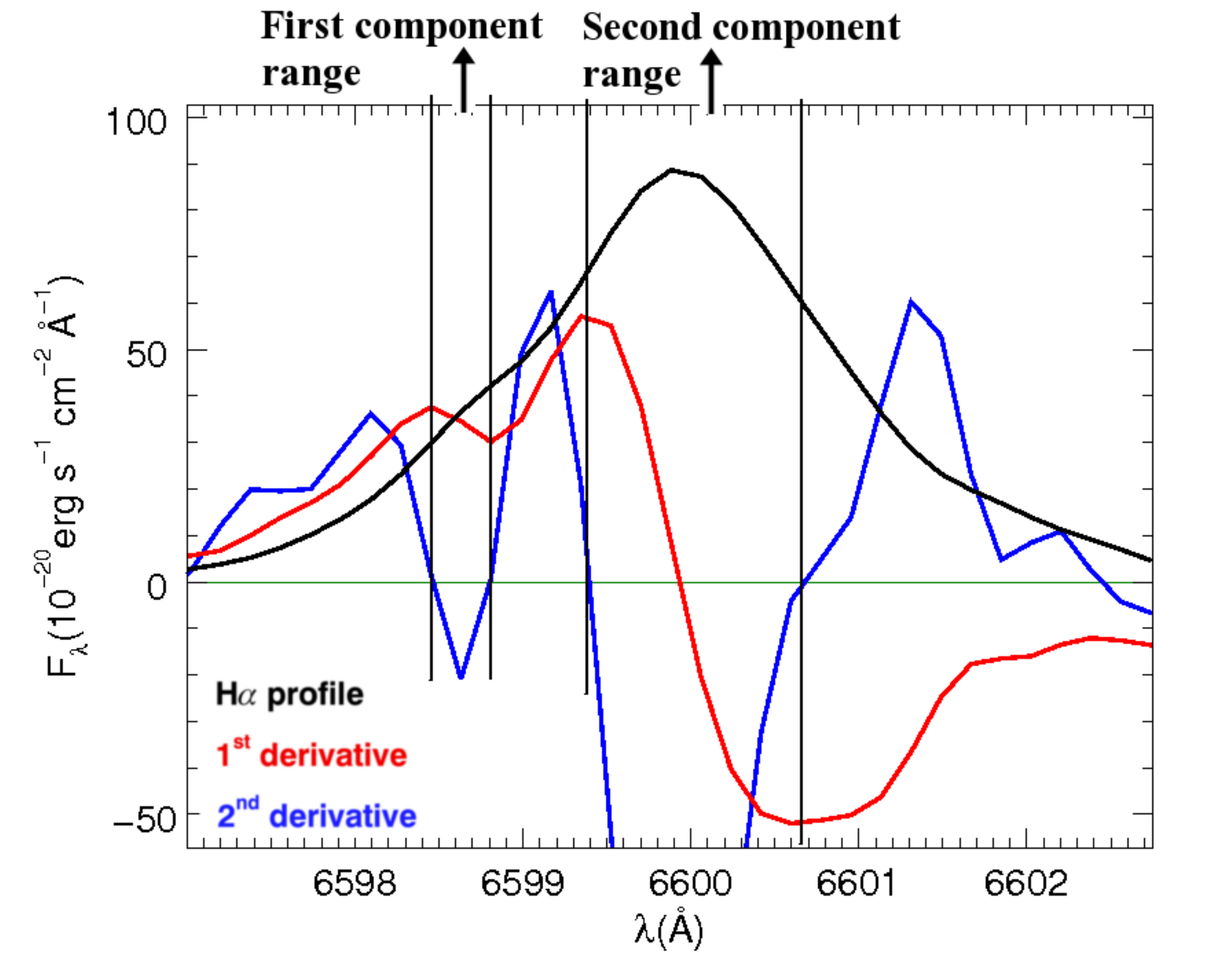}

\caption{H$\alpha$ profile (black), its first and second derivatives (red and blue, respectively). The vertical 
lines are the intersections of the maximum and minimum of the first derivatives and the zeros in the second 
derivatives.}
\label{fig:second_deriv_method}

\end{figure}

In order to effect best fits of multiple Gaussians to the profiles illustrated in Figure \ref{fig:second_deriv_method} 
we first approximated the parameters of the Gaussians, in order to restrict them to within the reasonable 
limits needed to obtain a convergent solution.

We have developed a method to find these approximations and 
their ranges using information derived from the first and second derivatives of the line profiles \citep{Goshtasby}. 

In Figure \ref{fig:second_deriv_method} we show profiles with structures whose basic shapes are repeated 
from positions over the whole circumnuclear zone, so that we can safely eliminate the possibility that they are noise. 
We assume that the line structures are composed of separate individual contributions which need to be separated. 
Although there are no well resolved peaks, we can clearly see the change in the profile gradients, which allows 
us to use these to effect the separation of the lines. For an emission profile with two intrinsic components, if we 
consider the profile behaviour as the wavelength $\lambda$ increases the centre of the first component will 
lie between the first maximum and the first minimum of the first derivative of the profile, \emph{i.e.} between the 
corresponding two zeros in the second derivative, (see Figure \ref{fig:second_deriv_method}). The centre of the second component 
will lie between the maximum and minimum of the first derivative in the other half of the profile, \emph{i.e.} between 
the corresponding zeros in the second derivative. This gives us a range of $\lambda$ for the centre of each of the 
two emission components, and the initial estimates of the line centres are made by taking the centre wavelengths of 
these ranges. Similarly the ranges give us initial estimates of the line widths, $\sigma_v$. We can then apply a 
standard Gaussian fitting routine with two components to a given observed profile, using the initial estimates as 
our starting points. 


 \begin{figure*}
\begin{tabular}{c|c}

\centering

\includegraphics[width=0.5\linewidth]{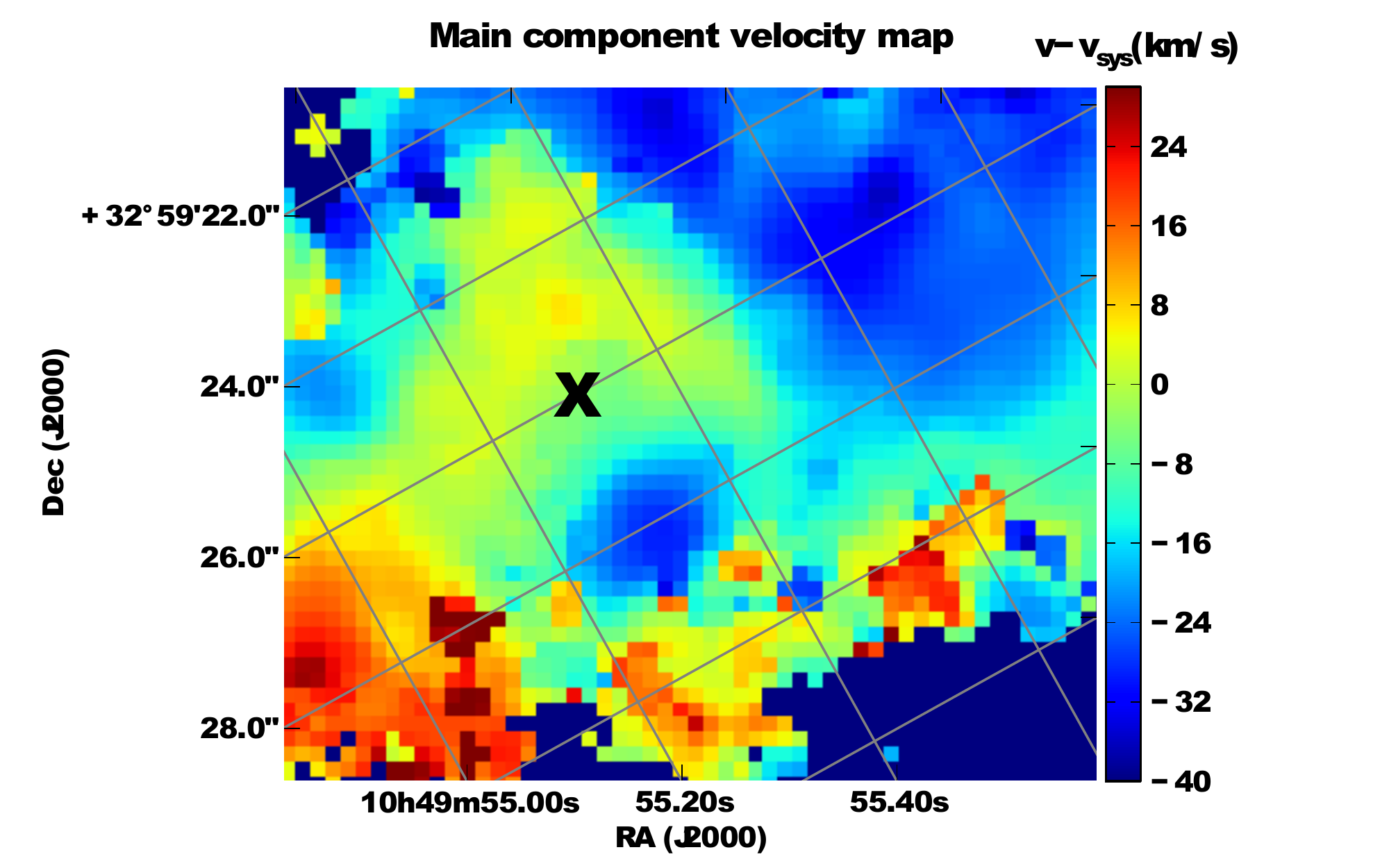} &

\includegraphics[width=0.5  \linewidth]{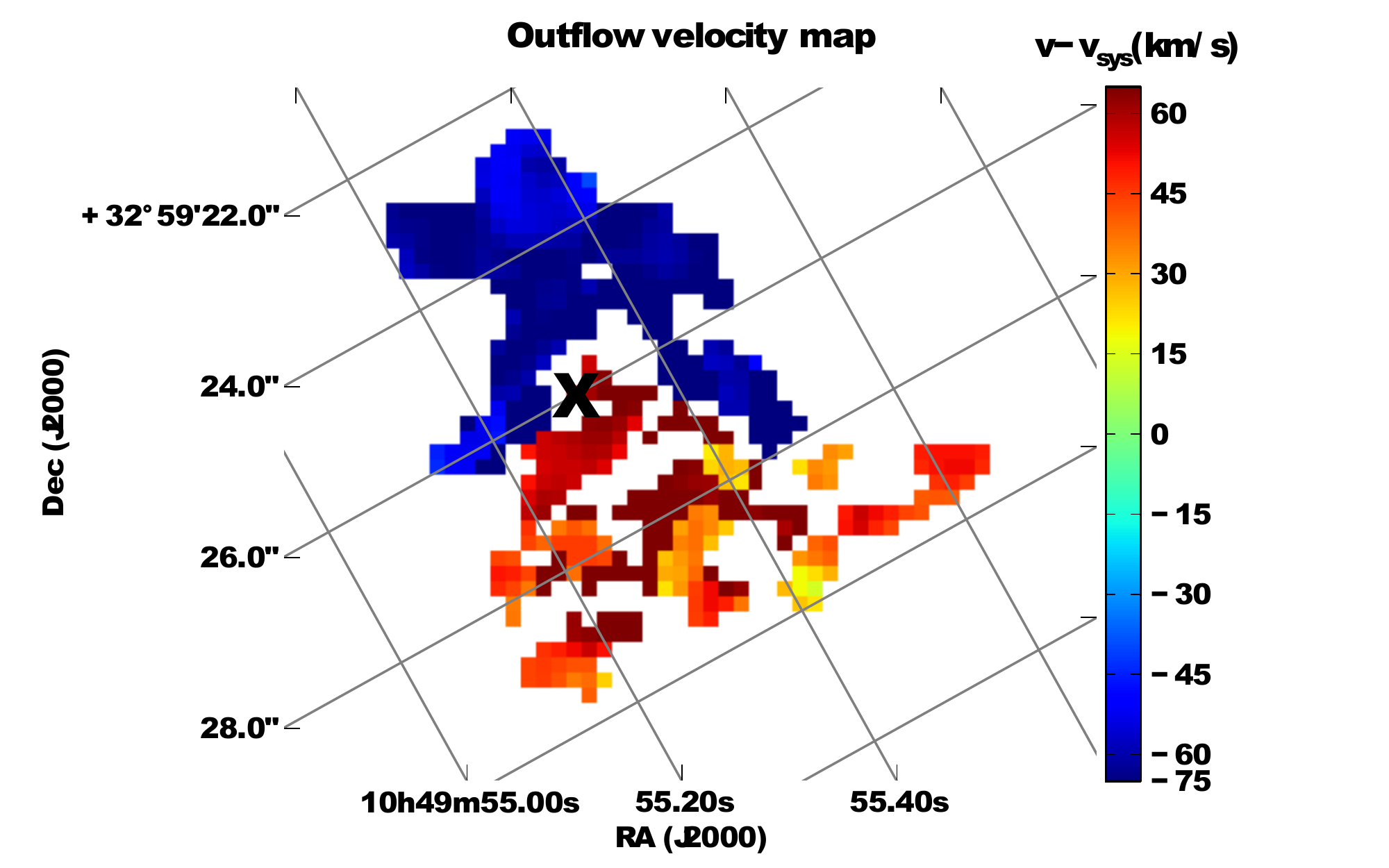} \\

\end{tabular}
\caption{Left: main component velocity map. Right: velocity map of the outflow. The ''X'' mark represents the 
kinematic nucleus.}
\label{fig:figure4}
\end{figure*}

The results of this method applied to the circumnuclear gas in NGC 3396 is shown in Figure \ref{fig:figure4}, where we show 
the velocity distributions of the components detected to either side of the nucleus, as seen clearly using 
the colour coding. These maps were made using the main (Figure \ref{fig:figure4} left)  
and the secondary (Figure \ref{fig:figure4} right) components of the line profiles. From the 
angular distribution of the secondary component (Figure \ref{fig:figure4} right), 
we argue that there are two independent velocity components around the nucleus very different 
from the typical inflows patterns (spirals and/or bars)
and from the velocity distribution we argue that the two velocity components are in opposite senses from the nucleus 
since the discontinuity in velocity is $~120 \, \mathrm{km/s}$ around the kinematical nucleus. We suggest that 
it represents a biconical outflow of ionized gas from the nucleus of NGC 3396 rather than an inflow prolongation. 
Using a method for estimating the mass outflow rate which is analogous to that described in Section $\S3.1$, 
we estimate the density as the electron density of the central region of $n_e=0.1\, \mathrm{e^{-}/cm^{-3}}$. We have 
used the $5\%$ of the $H\alpha$ luminosity in the central region since in the previous 
multiple components fit, we found that the outflow emits the $5\%$ of the total $H\alpha$ luminosity in the 
nucleus.

In order 
to estimate the section where the outflow occurs, we assume that the outflow is happening in a cylinder with radius 
equal to the radius of the base where the approaching and the receding part of the outflow join in Figure 
\ref{fig:figure4} (right), 
yielding a value of $R=(300\pm100))\, \mathrm{pc}$. We assume that the velocity of the outflow is the deprojected (perpendicular 
to the disc plane, $i=(67\pm 4)^{\circ}$) mean value of the 
absolute velocity values in Figure \ref{fig:figure4} (right), yielding $v_{out}=(150\pm 50)\, \mathrm{km/s}$

We derive a value 
for $\phi$ , the outflow rate, of $(0.22\pm 0.07) \, \mathrm{M_{\odot}/yr}$. 

The net difference between the inflow and outflow rates approximately coincides 
with the estimated SFR. 

\subsection {Interaction-induced nuclear activity?}

We think that the interaction has induced nuclear activity, in particular star formation and, maybe, a shrouded AGN.
The outflow is found in the NGC 3396 central ionized zone where $L_{H\alpha}=2.8\cdot10^{40}\, \mathrm{erg/s}$. The 
majority of the H$\alpha$ 
luminosity is due to a burst of star formation that we have estimated as $0.15\, \mathrm{M_{\odot}/yr}$ since 
we have found that the outflow emits only $5\%$ of the H$\alpha$ 
luminosity. The 
outflow is maybe due to the winds of the massive stars that are forming in the nucleus. However,   
although the morphology of the outflow is not precisely determined, the overall shape does seem to correspond to 
the biconical effect expected from an AGN, even though it is not classified as such. We have checked 
the possibility that there is an obscured AGN, finding that  
\cite{2005MNRAS.360..801B} did discover an X-ray source close to the H$\alpha$ nucleus, whose spectral 
hardness and luminosity could imply an origin in an AGN. 

\begin{figure}

\includegraphics[angle=-90,width=0.95\linewidth]{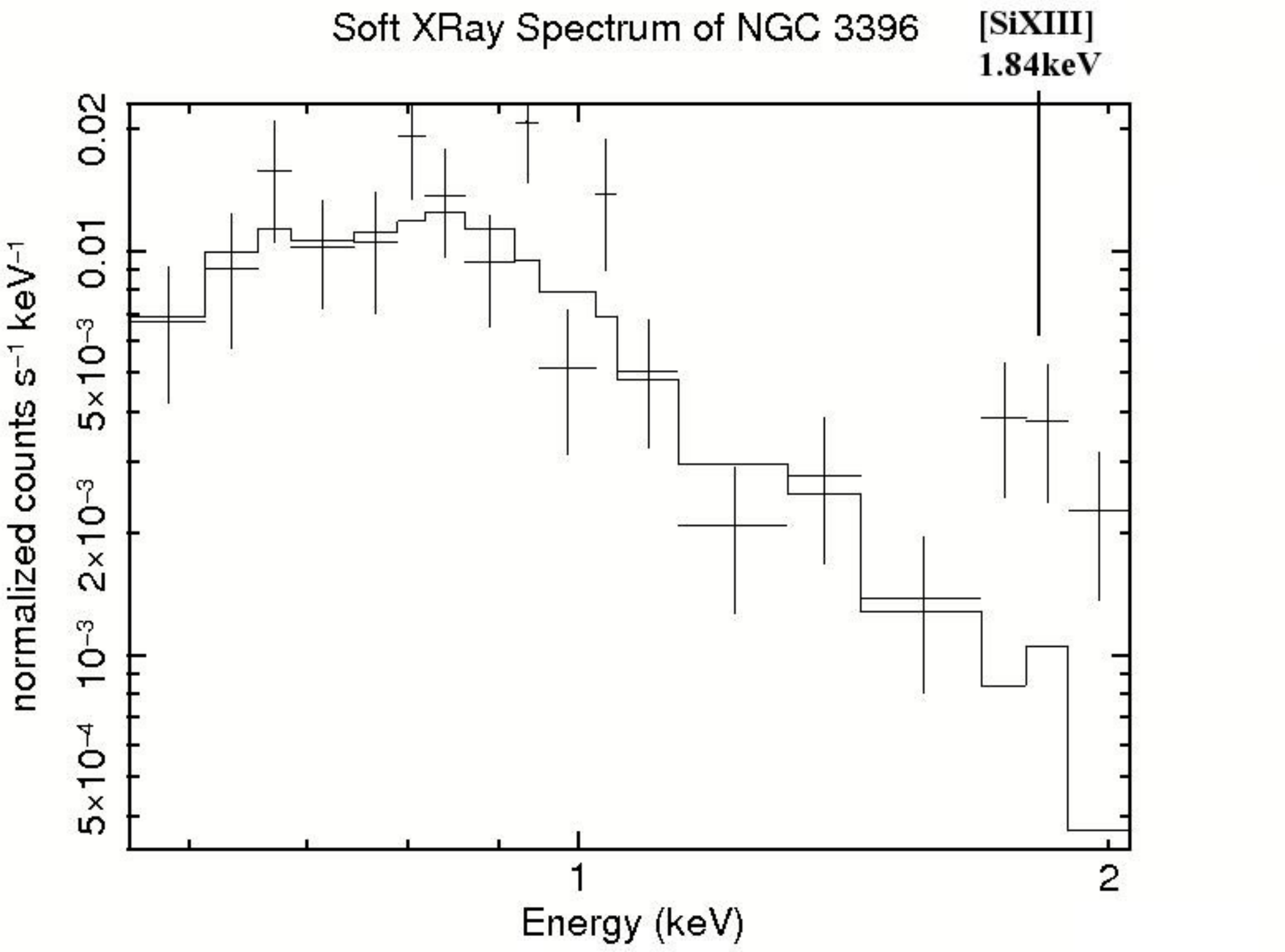}

\caption{Observed Soft X-Ray spectrum of NGC 3396 (points) and the VMEKAL model of the observed spectrum (continuous line).} 

\label{fig:figsx}
\end{figure}

We therefore extracted its X-ray spectrum from the Chandra data archive to look for indicators of the presence of 
AGN in this range. Unfortunately, the hard band (2-10 keV) presented very few counts and therefore we could 
not detect the [Fe K] lines which are the main indicator with acceptable certainty, though we did find an 
excessively prominent [Si XIII] line in the soft band (0.5-2 keV) at 1.84 keV (Figure \ref{fig:figsx}). 
The soft band was fitted with a line and continuum emission with tunable abundances model VMEKAL, initially 
assuming an abundance scheme derived from type II supernovae core collapse enrichment. This model failed to properly 
fit the [Si XIII] line by a factor of 
$\sim12$ in the model parameter, meaning the abundance needed to fit the line is $\sim12$ times 
that predicted by the SNII model. It has been found in a previous paper \citep{2011A&A...529A.106I}, 
in a sample of 44 galaxies, that this excess, of at least twice, shows a good correlation with the presence of an AGN.


\cite{2008ApJ...677..926S} argue that the presence of strong ionization lines such us 
[NeV] at $14.3\, \mathrm{\mu m}$ and $24.3\, \mathrm{\mu m}$, or maybe [OIV] at $26.0\, \mathrm{\mu m}$, is strong evidence for an obscured 
AGN. We have analyzed infrared spectra from Spitzer of the nucleus of NGC 3396 but we did not find 
any strong ionization line supporting the evidence for an AGN, so the question remains open.




The overall result of the use of kinematic evidence to study the bar and circumnuclear zone of NGC 3396 is, then, 
to attribute the underlying velocity field to gas inflow to the nucleus rather than to rotation, to discover evidence 
for biconical outflow from the nucleus, and to detect symptoms of nuclear activity, 
star formation and/or a shrouded AGN. 



\section{Physical properties of HII regions}

Using the H$\alpha$ emission from the two galaxies we can compare the SFR´s (estimated from 
\cite{2009ApJ...703.1672K} as in section $\S3.1$) in their circumnuclear zones 
with those in the external parts of the two galaxies. For NGC 3396 the circumnuclear SFR is $0.15\, \mathrm{M_{\odot}/yr}$, while 
in the rest of the disc the integrated SFR is $0.39\, \mathrm{M_{\odot}/yr}$. 
The corresponding 
figures for NGC 3395 are $0.08\, \mathrm{M_{\odot}/yr}$ 
and $1.1\, \mathrm{M_{\odot}/yr}$ 
respectively. Although we might 
think that the difference in the SFR´s between the 
discs and the centres could be because NGC 3396 is edge-on, this is not evident, as we have found that the kinematics 
of this object are not those characteristic of an edge-on disc. If NGC3396 were edge-on, it would be more 
affected by dust. However, 
we have seen in section $\S2$ NGC3395 is more affected by dust attenuation than NGC3396. 

In order to explore these different SFR's values in more depth we derived the directly observable physical parameters of 
the HII regions: the H$\alpha$ luminosity, $L_{H\alpha}$, the effective radius, $R$, and the  velocity dispersion, $\sigma_v$ , 
from the GH$\alpha$FaS data cube, using the clumpfind algorithm \citep{1994ApJ...428..693W}. We were 
careful to take into account the limitations of clumpfind \citep{2009ApJ...699L.134P}, estimating the 
completeness limits in order to obtain results unbiased by sensitivity and resolution considerations.  
We used a lower limit of 9 for the number of pixels per statistically significant region, and eliminated pixels 
from the regions selected which had S/N ratios less than 2. These two criteria set a lower limit for the luminosity of  
$L_{H\alpha\thinspace min} = 18 \, \mathrm{rms} = 2\cdot \, \mathrm{rms}\cdot \pi \cdot R_{min}^2$, where rms is the rms noise level in the background 
of the H$\alpha$ signal. Another lower limit we have used is a lower limit in 
the non-thermal velocity dispersion, $\sigma_{v-nt}$, in $8\, \mathrm{km/s}$, the uncertainty in our velocity measurements, 
since we are going to use it to estimated the virial mass. We described the 
non-thermal velocity dispersion estimation bellow (section $\S4.2$).

We made statistically valid measurements on 108 HII regions, 20 of which belong to NGC3996 and 88 to NGC3395.

We found results comparable to those obtained by \cite{2011AJ....141..113G} who used narrow band imaging 
observations in H$\alpha$ surface brightness to determine the radii and luminosities of HII regions in M51. 
The additional dimension of velocity we use here serves to yield less scatter in the physical
properties of the regions, as it reduces significantly the effect of confusion in the sources.
Using the simplifying assumption of sphericity for the regions, we can obtain
the mean electron density $n_e$ (in fact the luminosity-weighted electron density $<n_e>$, equation \ref{eq:density}),
and hence estimate the ionized hydrogen
mass ($M_{HII}$) of each region. In section $\S4.4$ we will describe how we have estimated the
fraction of the ionized
gas compared with the neutral atomic plus the molecular gas, and why it seems approximately constant, at least for the regions
with $L_{H\alpha}>10^{38.7}\, \mathrm{erg/s}$. Using the ionization fraction of $0.2$ we have then estimated the total
gas mass, $M_{gas}$ (ionized + neutral + molecular).

We also derived the virial mass of each region
following \cite{2012ApJ...750..136W}, using the expression they used assuming self-gravitating region in virial
equilibrium,

\begin{equation}
 M_{vir} = \frac{3 \thinspace R \thinspace\sigma_{v-nt}^2}{G}
\label{equation:virial}
\end{equation}

where $R$ is radius of the region we have measured. 


We present the results of the physical properties in Tables \ref{table:prop1}, \ref{table:prop2} and \ref{table:prop3}

\subsection{Parameter analysis}

\begin{figure*}
\centering

\includegraphics[width=0.95\linewidth]{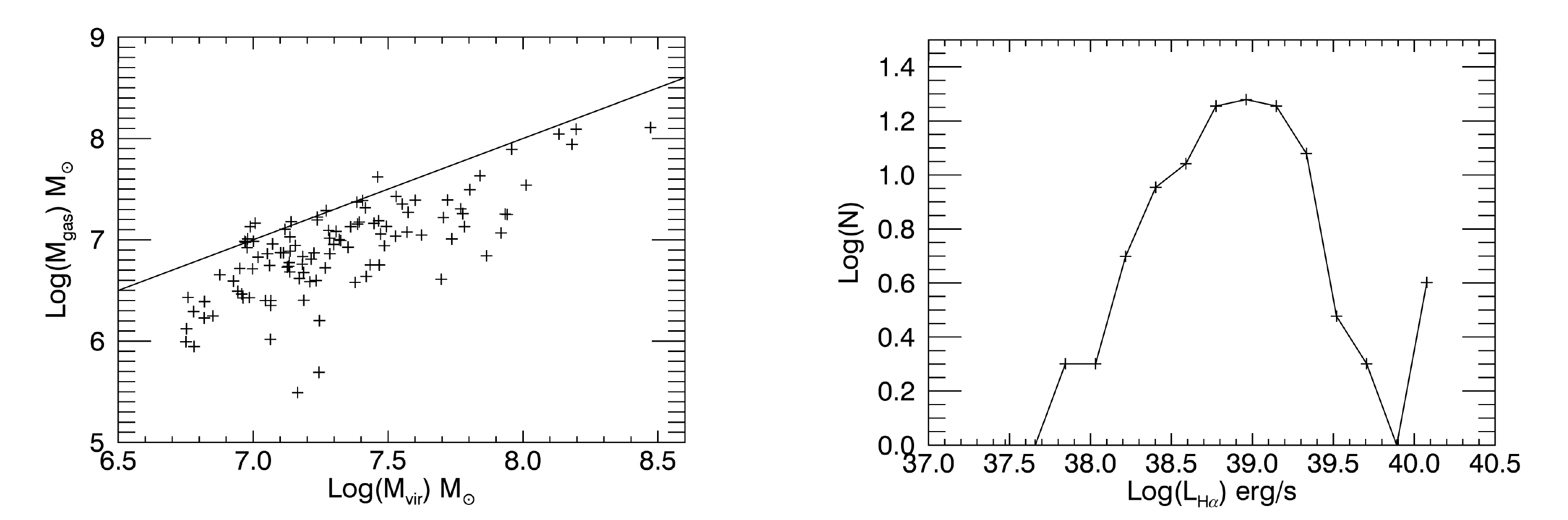}

\caption{Left: Comparison between gas mass and virial mass. Right: luminosity function of the HII regions}
\label{fig:figure9}

\end{figure*}

Figure \ref{fig:figure9} (left panel) presents the comparison of $M_{vir}$ and $M_{gas}$, showing 
that the gas masses of the regions are generally lower than the virial masses, 
although there is a clear correlation between them. In Figure \ref{fig:figure9} (right panel) we show the luminosity 
function of the HII regions. There is an apparent break in the function at 38.6 dex, although 
we cannot claim accurate information here, because it is not easy to specify the lower completeness limit. We note 
that a break at this luminosity was first detected by \cite{1989ApJ...337..761K}; a more detailed 
study involving 57 galaxies was produced by \cite{2006A&A...459L..13B}, while 
\cite{2000AJ....119.2728B} claimed that this break was due to a change from 
luminosity-bounded (in the lower luminosity regime) to a density-bounded regime (in the higher 
luminosity regime). However here we have the extra parameter of 
velocity dispersion which can give us further insights.

\subsection{Two populations of HII regions}

\begin{figure*}
\centering

\includegraphics[width=0.95\linewidth]{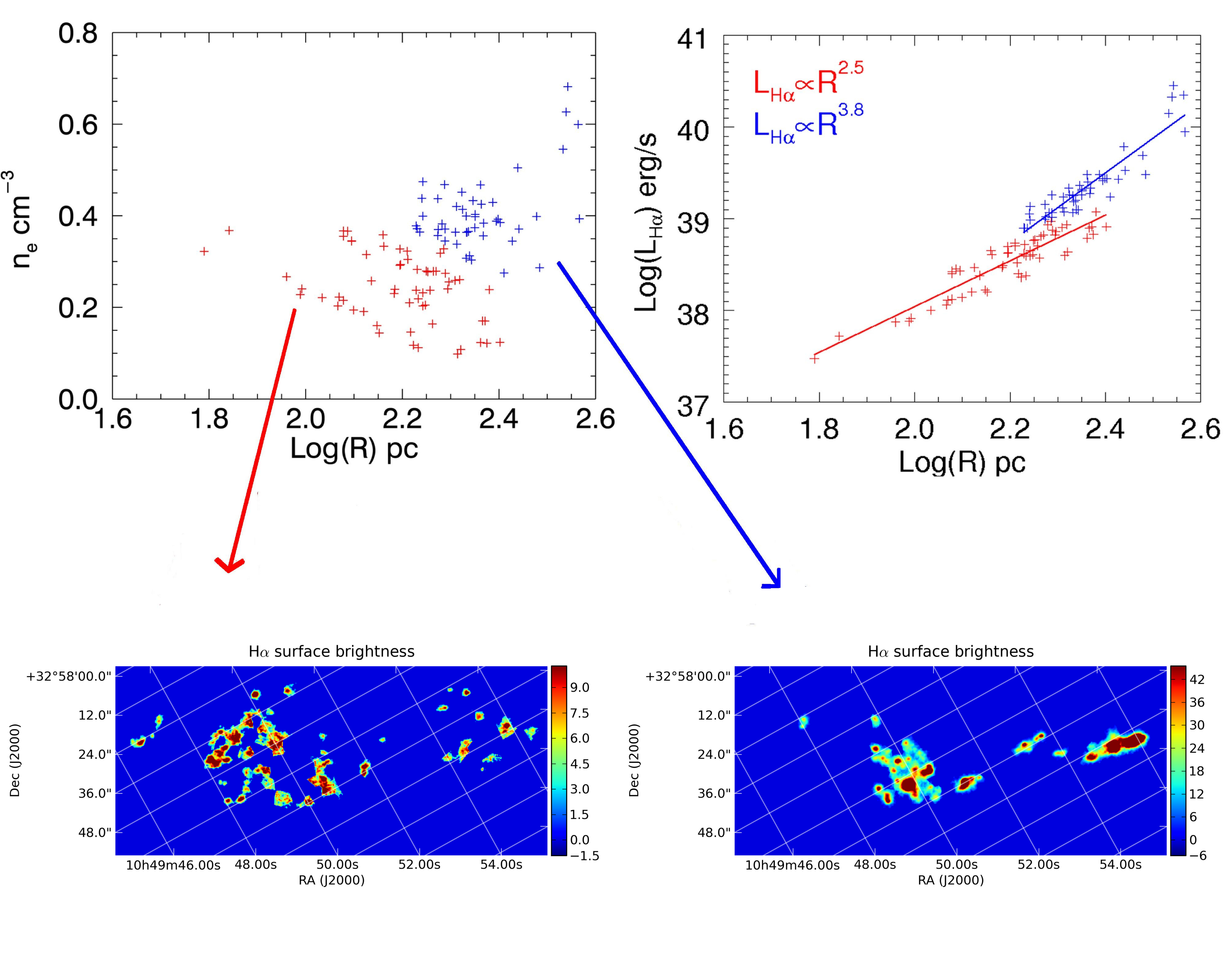} 




\caption{Relation between electron density, $n_e$, and HII region effective radius, $R$, (top-left). 
Relation between H$\alpha$ luminosity, $L_{H\alpha}$, and HII region effective radius, $R$ (Top-right). 
Red points are for the the decreasing $n_e$ regime and 
the red ones for the increasing one. Lower luminosity (bottom-left) and 
higher luminosity (bottom-right) HII regions.}
\label{fig:figure5}

\end{figure*}

A first step in our analysis involved plotting the mean electron density $n_e$ against the radius $R$ of the region,
and also the H$\alpha$ luminosity, $L_{H\alpha}$ against $R$, as shown in the upper panels of Figure \ref{fig:figure5}.
Although it is possible to fit a non-linear curve with continuous change in slope to the latter, it is instructive
to use the double linear fit shown in Figure \ref{fig:figure5} (upper right panel).

The selection of the points as ``blue`` or ``red`` was made using a division into two groups
The separation between the groups is clearer in
the electron density-radius plot than in the luminosity-radius plot since
there are regions where the electron density falls with radius, and there are regions where the
electron density rises with radius,  (Figure \ref{fig:figure5} upper left)
Thus, in order to chose the correct separation we have
assumed that the two regimes are separated in electron density-radius space. It is possible that the boundary is 
less well-defined than we have imposed here,
but we are not as concerned to find a precise boundary between the two populations as we are to study the collective
properties of the two different populations, and there may be a minor degree of mixing between the two 
populations near the boundary. We will show below, using the geometrical distribution of the two sets of regions, 
why this is in practice unlikely.
We have chosen the separation of the two populations by means of the $\chi^2$ minimization
(in the electron density-radius space) and a double linear fit for several possible separations.
The result is the less luminous set of HII regions show
an index of 2.5, while the more luminous set shows an index of 3.8.

We find that there are 9 regions in the lower luminosity regime and 11 in the higher
luminosity regime in NGC3396, and for NGC3395 the corresponding numbers are 49 regions in the lower luminosity regime 
and 39 in the higher luminosity one. The higher luminosity regions in NGC3396 emit 94\% of the H$\alpha$ luminosity while 
for NGC3395 these regions emit 75\%. If we scale the  SFR as proportional to the H$\alpha$ luminosity (this is the only 
SFR tracer used here ) this would imply that
the higher luminosity regime represents 94\% of the SFR in NGC3396 and 75\% of the SFR in  NGC3395.

We have distinguished between these two regimes because the $n_e$ decreasing regime 
agrees qualitatively with the result 
reported in \cite{2010ApJ...710L..44G} who found, in M51 (for over 2000 HII regions) and in 
NGC 4449 (for over 200 regions), that $n_e$ varies as $\frac{1}{R^{0.5}}$. In their samples there were very 
few regions with luminosities higher than 38.6dex so they were essentially looking at the equivalent of the lower 
luminosity set of regions we are observing in Arp 270. This is due to the intrinsic properties of 
the luminosity functions in the two galaxies they observed, which means that the upper luminosity population 
we observe here is not represented in their work. In Figure \ref{fig:figure5} (upper left panel) we see  that there is a 
tendency of the high luminosity regions, in blue, to show an increase in $n_e$ with increasing radius, the opposite tendency, 
at least in qualitative terms, from that exhibited by the lower luminosity set. 
It is worth noting two additional points here. Firstly \cite{2010ApJ...710L..44G} found not only the $\frac{1}{R^{0.5}}$ 
dependence mentioned above, but also a very clear dependence of $n_e$ on the galactocentric radius of the HII region, 
which they interpreted as showing that the regions in their sample were pressure bounded. We will return to this point later, 
but it is mentioned here because we are not in a position to include this dependence in our results. 
Not only do the distance and the orientations of the two galaxies make any estimate of galactocentric radius very 
inaccurate, but the interaction itself will certainly affect the pressure distribution which confines 
the HII regions. 
This is at least one of the reasons why the dependence of $n_e$ on radius shows so much scatter in Figure \ref{fig:figure5}. 

A second point is brought out in the lower panels of Figure \ref{fig:figure5} where we show the H$\alpha$ surface brightness
of the individually reconstructed map for each individual region, indicating the locations within
the galaxies of the HII regions whose parameters are plotted in the upper panels. All the regions plotted as
blue points in the upper panels are found in concentrated zones around the centres of the two galaxies, and in the
interaction zone. All the regions plotted as red points are found in more peripheral zones of the galaxies. The zones
in the two lower panels are precisely complementary, and do not overlap. This purely geometrical
separation could have been used to distinguish the two populations in the first place, and it can be used to 
give good support to the separation criterion we in fact used, based on the radius-electron density diagram.

The general dependence of the H$\alpha$ luminosity on the parameters of an HII region is analyzed via equation \ref{eq:density}
where we neglected, for simplicity, the presence of helium and other elements, and if the hydrogen is fully ionized, 
equation \ref{eq:density} implies that the luminosity should depend simply on $R^3$ in case of 
$n_e$ were not R dependent. We have seen from Figure \ref{fig:figure5} 
that this relation does not hold, either for the lower luminosity regime, where the exponent is 2.5, or for the 
higher luminosity regime, where the exponent is 3.8. In the former case we can infer that $n_e$ must be a declining 
function of R, varying as $\frac{1}{R^{0.25}}$. This is not in good quantitative agreement with the results 
of \cite{2010ApJ...710L..44G}, but as explained above, we have not been able to disentangle the macroscopic effects 
of pressure variations within the galaxies for Arp 270. In the higher luminosity range, $n_e$ must vary as $R^{0.4}$, 
and this can be explained if the effect is due to the increasing ionization of the higher density clumps, 
as outlined above.

\begin{figure*}
\centering

\includegraphics[width=0.95\linewidth]{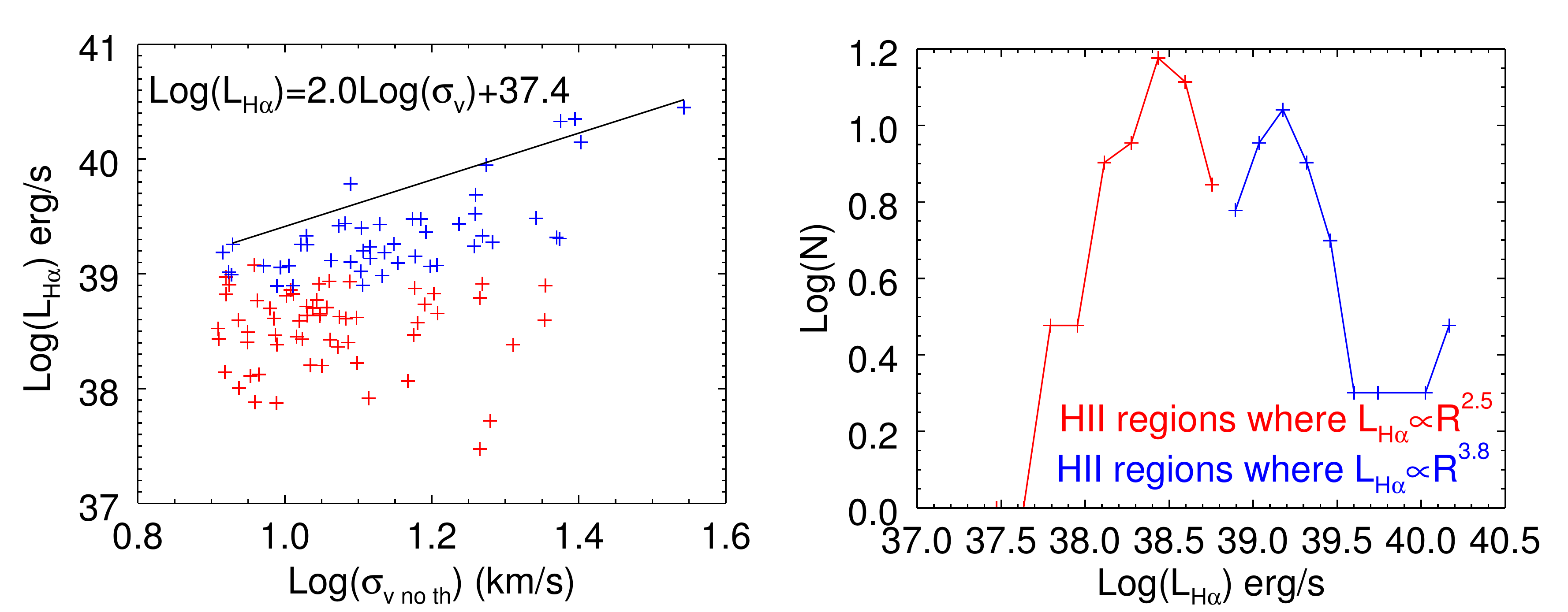}

\caption{Relation between the H$\alpha$ luminosity, $L_{H\alpha}$, and the non thermal velocity 
dispersion, $\sigma_{v-nt}$, for each individual HII region. We have overlapped the fitted envelope,
estimated following \citet{2005A&A...431..235R} (left). Luminosity function (right)}
\label{fig:figure7}

\end{figure*}

In Figure \ref{fig:figure7} (left hand panel), we have plotted $L_{H\alpha}$ against the non-thermal velocity dispersion, 
obtained from the observed velocity dispersion by subtraction in quadrature of the instrumental, thermal, and natural 
line widths. We have considered that the quadratic subtraction overestimates 
the non-thermal velocity dispersion in $~8\, \mathrm{km s^{-1}}$ \citep{2010MNRAS.407.2519B}, so we have subtracted it. 

$L-\sigma$ relation for HII regions was first looked at in detail by \cite{1981MNRAS.195..839T}. They 
found that it could be expressed as a power law, $L\propto \sigma^N$, and suggested that N should take the value 4 for 
virialized regions. However in practice they, and subsequent studies, found values $N\in[1.0-6.6]$. 
Here we follow the method of \cite{2005A&A...431..235R} who suggested that points on the upper envelope in $L$ 
(the lower evelope in $\sigma$) should represent regions in virial equilibrium. Our result for 
this envelope is $Log(L_{H\alpha})=2.0\thinspace Log(\sigma_v)+37.4$, which as we can see has a slope of 2 in 
Figure \ref{fig:figure7} left. In the last approach, the high luminosity regions tend to be more virialized 
(blue points in Figure \ref{fig:figure7} left) than the low luminosity regions (red points in Figure \ref{fig:figure7} left).


It is interesting to note how, as shown in Figure \ref{fig:figure7} (right hand panel), plotting the 
luminosity function separately for the HII regions in the two regimes gives an important clue to 
the origin of the discontinuity found in the undifferentiated luminosity function at $L_{H\alpha} =38.6\, \mathrm{dex}$. 

\begin{figure}
\centering

\includegraphics[width=0.98\linewidth]{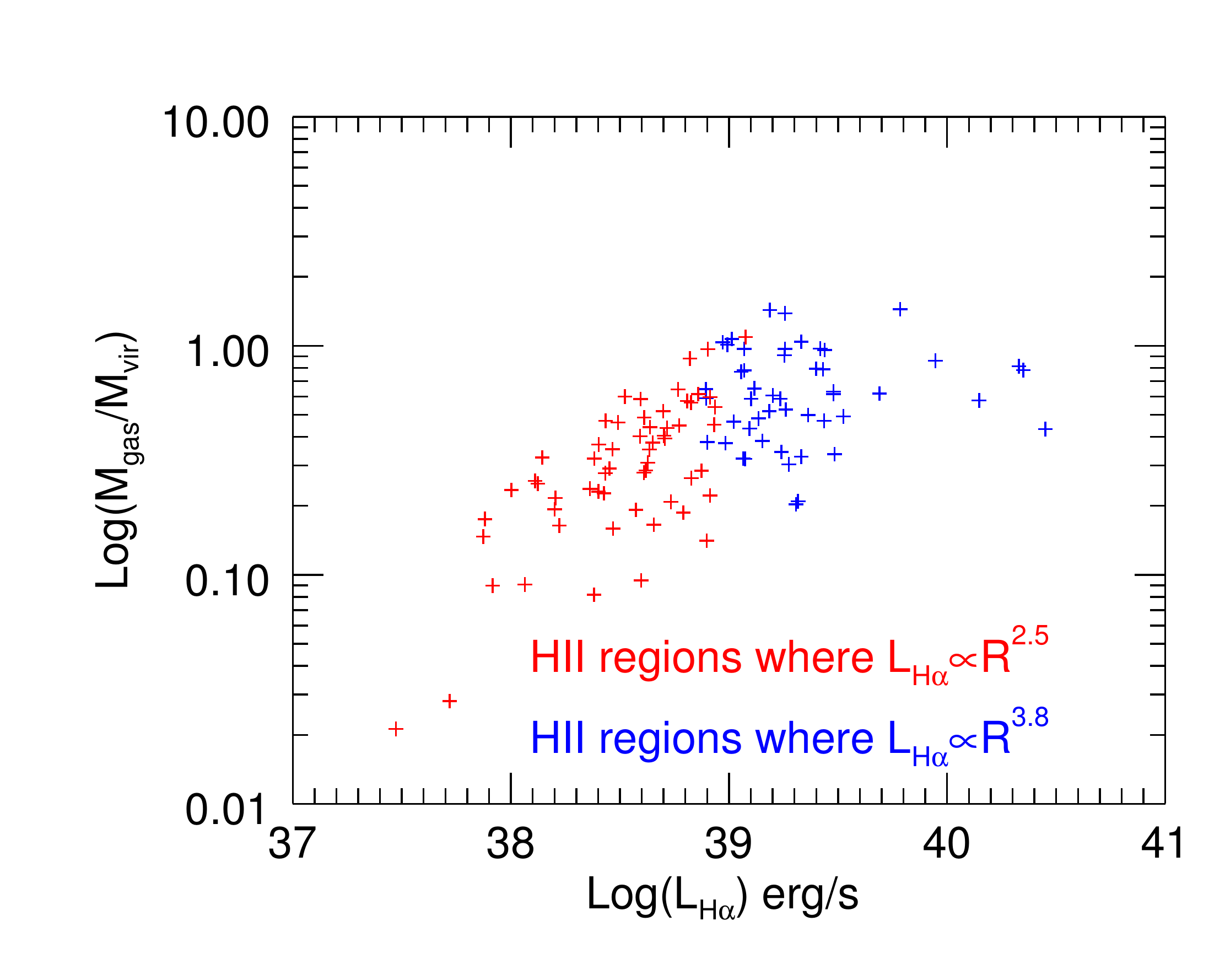}


\caption{Total gas mass $M_{gas}$ over virial mass $M_{vir}$ versus $H\alpha$ luminosity for the higher
luminosity regime (blue crosses) and lower
luminosity regime (red crosses).} 
\label{fig:fig_xfrac}

\end{figure}

In Figure \ref{fig:fig_xfrac} we have plotted the ratio of the total gas mass to the virial mass.
We note the tendency of this ratio to rise for larger radii, and higher masses for the regions in the
lower luminosity regime. This is presumably
because of the combination of two effects.

Firstly, the estimate of the total gas mass  depends on the ionized gas mass due to the
method we have used (described in $\S4.4$) in estimating the total gas content for the lower
luminosity regime. We argue that
an increasing fraction of the hydrogen is ionized with increasing luminosity in this regime.
Secondly, these regions are not virialized since they are bellow the envelope in the
L-$\sigma$ relation (Figure \ref{fig:figure7} left), so the derived virial mass is in practice an overestimate of the
gas mass. 



However, there is a tendency for the gas mass to approach the virial mass for the higher luminosity regime.
We can explain these observations qualitatively again, with a virialized scenario for HII regions for the higher luminosity
regime since their gas masses approach parity with the virial mass.




\subsection{Differences between the two regimes}

\cite{2010ApJ...710L..44G} showed that for the HII regions in M51 and NGC 4449 the decline in their values of 
$n_e$ with galactocentric radius, with scale lengths equal to the scale lengths of the HI surface 
densities in the two cases. This implies that the regions are in pressure equilibrium with the surrounding gas in 
the discs, and are in consequence pressure bounded. However, in \cite{2011AJ....141..113G} it was noted that in both 
of these galaxies the number of regions with luminosities greater than $L_{H\alpha} = 38\, \mathrm{dex}$ is small, with a 
complete absence of regions with luminosities greater than 39 dex. The difference between Arp 270 and these other 
objects seems to be that the interaction has stimulated the production of HII regions with luminosities higher than 39 dex, 
which are those in the high luminosity regime (regime 2) in the present article. So how can we account for the 
differences between the two regimes.
One clue can be found in two articles dealing with another well-known pair of interacting galaxies: the Antennae. 
In \cite{2012ApJ...750..136W} and in \cite{2013font} the authors detect two populations of giant molecular clouds, 
with a division between them at a cloud mass of $10^{6.5}\, \mathrm{M_{\odot}}$.
This is close to the value of the gas mass which marks the change from regime 1 to regime 2 in Arp 270, and 
\cite{2013font} have detected two equivalent regimes in the HII regions of the Antennae. These results lead 
us to postulate that the HII regions in regime 2 (high luminosity) are those which have
 formed in  molecular clouds of high mass, which have own ''regime 2``, while the HII regions in regime 1 
(low luminosity) have formed 
from molecular clouds in its ''regime 1``. 
We argue that while the HII regions in regime 1 are pressure bounded, those in regime 2 are gravitationally bound, 
(and bounded).  Those with virial masses close to the upper envelope of the plot in Figure \ref{fig:figure9} 
(left panel) we will take as in virial equilibrium, and therefore gravitationally bound implying that the 
line widths of those regions are largely due to gravitation. 

Further evidence supporting the idea of gravitationally bound regions can be found by making an estimate
of the Jeans mass ($M_J$) from the equation (5.47) of \ref{2008gady.book.....B},

\begin{equation}
 M_J=2.92\frac{v_s^3}{G^{3/2}\rho_0^{1/2}}
\end{equation}

where $v_s$ is the speed of sound and $\rho_0$ is the density. We have assumed $v_s=\sigma_{v-nt}$ and
$\rho_0= \frac{M_{gas}}{(4/3)\pi R^3}$.

The comparison between the
Jeans mass and the total gas mass of each region yields a relation very similar to the plot of the comparison between
total gas mass and virial mass (Figure \ref{fig:fig_xfrac}), implying that for the higher luminosity regions
gravity dominates the equilibrium ($M_{gas}\simeq M_J$)
while the lower liminosity regions appear to be held in quasi-equilibrium by the external pressure of 
the surrounding gas column, given that $M_{gas}<M_J$ (see also \cite{2010ApJ...710L..44G} who 
find a different argument for pressure equilibrium in the HII regions of M51, which fall in the low luminosity regime). 

The turbulence within 
the regions is assumed to maintain the regions against collapse (and, more surprisingly, but factually, the same 
is true of the molecular clouds in the Antennae, where the turbulent line widths are large enough to give 
support against gravity). 

\begin{figure}
\centering

\includegraphics[width=0.9\linewidth]{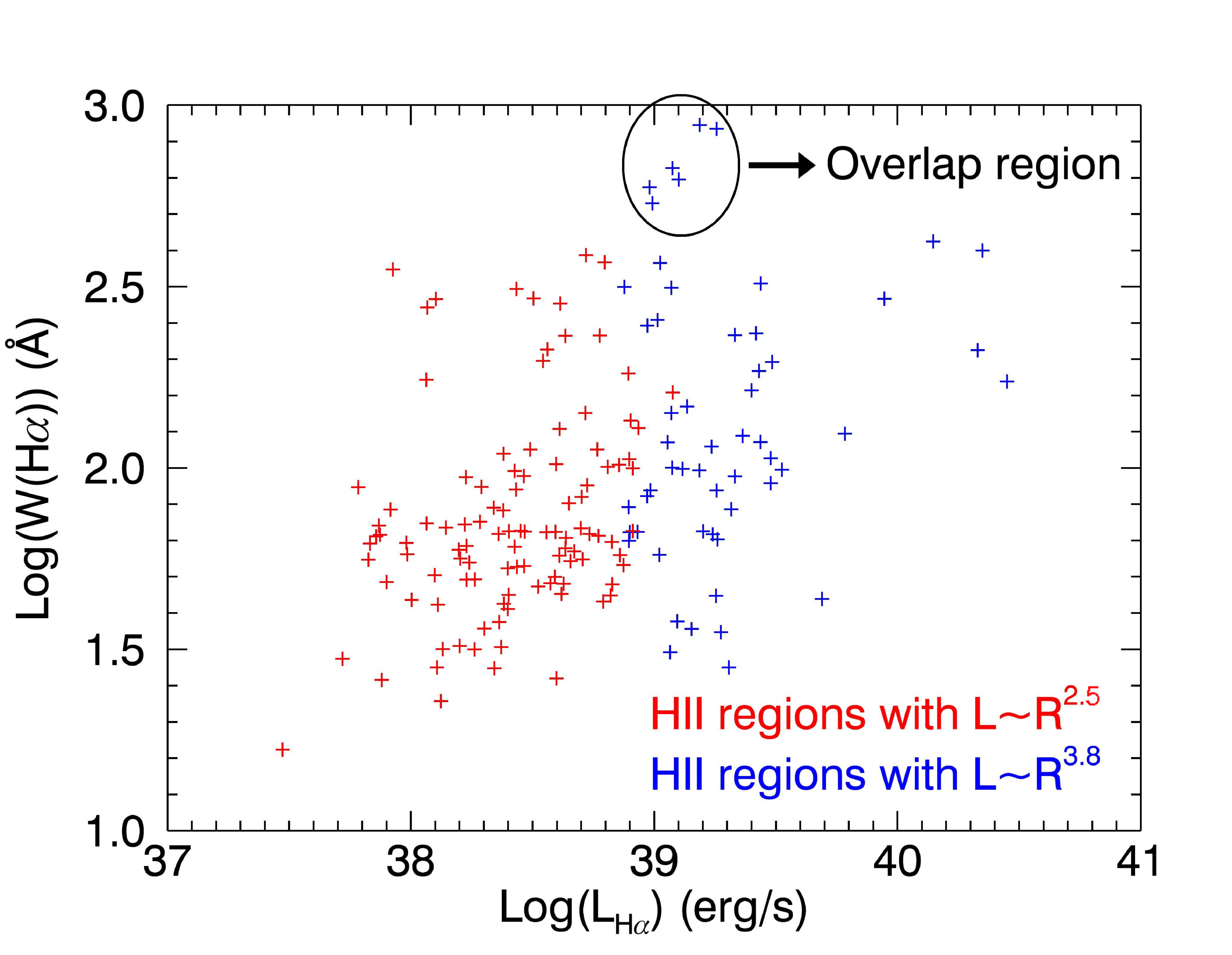}

\caption{$H\alpha$ equivalent width ($W(H\alpha)$) versus $H\alpha$ luminosity ($L_{H\alpha}$)}
\label{fig:fig_eqw}

\end{figure}

Another possible cause of the two regimes could be the ages of the regions. As the star formation 
in galaxy mergers tends to occur in the central regions of the galaxies, one might argue that 
the difference in the regimes is due to a difference stages in the HII regions evolution in time. 
We have plotted the $H\alpha$ equivalent width versus $H\alpha$ luminosity for each HII region in Figure
\ref{fig:fig_eqw}. Assuming a direct relation between $W(H\alpha)$ and the region age \citep{1999ApJS..123....3L}, 
we can conclude from Figure \ref{fig:fig_eqw} that the difference between regimes does not seem to be related to 
the region's ages. However, what we found in Figure \ref{fig:fig_eqw} is that the youngest regions 
($Log(W(H\alpha))>2.7$) belong to the overlap region between the two galaxies.

We can add, somewhat speculatively, that if gravity is sufficiently dominant, the cloud 
could have a negative heat capacity \citep{1968MNRAS.138..495L}, so that a rise in temperature would cause it to 
contract. Under these conditions the injection of photoionization energy could cause a rise in density 
leading to enhanced star formation. Thus the two regimes may arise because of the differences between 
pressure bounded and gravity bounded molecular clouds, which in turn lead to differences in the dominant modes 
of star formation within them.

\subsection{Results for global star formation.}

Our observations allow us to show how the star formation rate surface density depends on the 
surface density of the gas. 
We can assume from the measurements we have made, that $L_{H\alpha} \propto R^N$ with $N=3.8$ 
is a good approximation for the high luminosity regions. Then, the 
mean electron density $n_e$ varies as $R^{\frac{N-3}{2}}$, obtaining 
expressions for the SFR surface density, $\Sigma_{SFR}$, and the ionized gas surface density $\Sigma_{ion \thinspace gas}$ 
in equation \ref{eq:2} and equation \ref{eq:3} respectively.


\begin{equation}
 \Sigma_{SFR}\propto \frac{L_{H\alpha}}{R^2}\propto R^{N-2}
\label{eq:2}
\end{equation}
\begin{equation}
 \Sigma_{ion\thinspace gas}\propto n_e \cdot R\propto R^{\frac{N-1}{2}}
\label{eq:3}
\end{equation}

\begin{figure}
\centering

\includegraphics[width=0.9\linewidth]{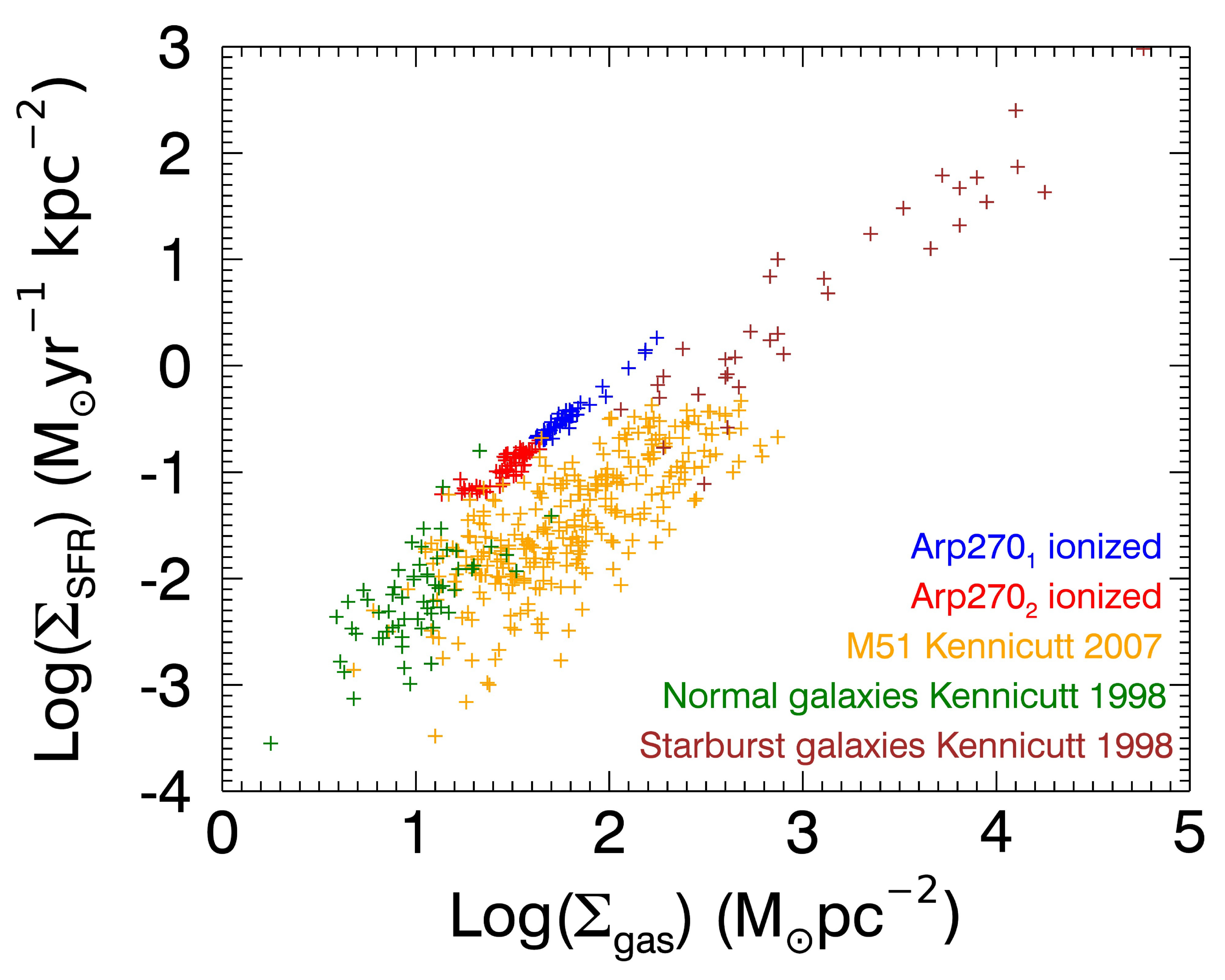}

\caption{SFR Surface density versus (ionized in the case of Arp 270) gas surface density}
\label{fig:figure8}

\end{figure}

Combining equations \ref{eq:2} and equation \ref{eq:3} we have 
\begin{equation}
\Sigma_{SFR}\propto \Sigma_{ion\thinspace gas}^{\frac{2N-4}{N-1}}
\label{eq:4_1}
\end{equation}

and inserting the observed value of 3.8 for N we find that 

\begin{equation}
 N=3.8 \rightarrow  \Sigma_{SFR}\propto \Sigma_{ion\thinspace gas}^{1.3}
\label{eq:4}
\end{equation}

which shows that the HII regions in the high luminosity regime do obey the Kennicutt-Schmidt law for 
global star formation \citep{1998ApJ...498..541K}. 
We can see this in Figure \ref{fig:figure8}, where we show the
values of the SFR surface density versus (ionized, in the case of Arp 270) gas surface density. In order
to compare with the Kennicutt-Schmidt law, we show the values of normal galaxies, starburst galaxies
taken from \cite{1998ApJ...498..541K} and the values of the M51 regions taken from \cite{2007ApJ...671..333K}.

The offset for the points from Arp 270 is due to the absence of measurements of the neutral and molecular gas components in
the system. If we multiply per 5 the mass of the ionized gas, the offset would disappear, so we estimate the fraction
of the ionized gas respect to the neutral plus molecular gas as $\sim0.2$. This is the value we used to
estimate the total gas mass, in section $\S4$. 

\section{Summary \& Conclusions}
We have shown that there is gas inflow along the bar of NGC 3396 at a rate of $(0.40\pm0.2)\, \mathrm{M_{\odot}/yr}$, sufficient to maintain 
the observed star formation rate of $0.15\, \mathrm{M_{\odot}/yr}$.  We have also detected a biconical gas outflow 
from the nucleus at a rate of $(0.22\pm0.07)\, \mathrm{M_{\odot}/yr}$, and with velocities $150\pm 50\, \mathrm{km/s}$. 
If this outflow 
is generated by energy and momentum input from matter falling onto a supermassive nuclear black hole, 
it would take an inflow onto the black hole of only $4\cdot10^{-5})\, \mathrm{M_{\odot}/yr}$ converted into kinetic energy 
and outflow momentum with an efficiency of 10\% in order to fuel this superwind, so the hypothesis that the 
outflow detected indicates the presence of an AGN is relatively easy to sustain. 
We claim 
that this may show the presence of an interaction-induced nuclear activity, in the sense that the substantial mass inflow rate 
along the bar gives a straightforward mechanism for feeding the possible nuclear black hole, or the burst of star formation. 

We have found two sets of HII regions in Arp 270, with a dividing line in H$\alpha$ luminosity at 38.7 dex. Those with 
high luminosity (regime 2) are found in zones closer to the centres of both galaxies, and in a specific 
interaction zone (the overlap region), while those in the lower luminosity regime (1) are found at larger galactocentric radii 
in the relatively less disturbed parts of the galaxy discs. 
The relations between the mean electron density $n_e$ and the radius of the region differ in the two regimes.
We use arguments based on previous work to show that the regions in regime 1 are in pressure
equilibrium with their surroundings, and argue that those in regime 2 are very probably gravitationally bound noting
that the higher luminosity regions show especially high velocity dispersion values.
This would favor a gravitationally triggered star formation scenario.
The higher luminosity regime represents 94\% of the SFR in NGC 3396, and 75\% of the SFR in NGC 3395. 
The presence of a statistically significant 
sample of HII regions in regime 2 in NGC 3396 is surely due to the interaction between the galaxies, which has brought about 
inflow of gas towards their centers, and enhanced the density there and in limited regions of the discs which 
are in direct collision. However, for the existence of two star forming 
regimes of the kind we detect here galaxy galaxy interaction may well not be necessary. 
The evidence cited in the introduction 
for a change in slope of the luminosity function at $L_{H\alpha} = 38.6\, \mathrm{dex}$ has been found previously in many galaxies, 
and the only condition appears to be the necessity for a sufficiently large number of HII regions at 
higher luminosities to be able to make statistically meaningful measurements of this function, and of the 
parameters we have measured in the present article, namely $n_e$, $\sigma_{v-nt}$, and $R$. The two basic 
conditions needed for this are either that the galaxy itself be sufficiently massive and gas rich to 
produce the appropriate population of placental molecular clouds, or that massive molecular clouds are 
produced during an interaction.
 
\section*{Acknowledgments}
Based on observations from the William Herschel Telescope, 
Isaac Newton Group of Telescopes, Observatorio del Roque de los Muchachos, 
Insituto de Astrof\'isica de Canarias, La Palma, Spain. Support was 
from projects AYA2007-67625-CO2-01, AYA2009-12903 and AYA2012-39408-C02-02 
of the Spanish Ministry of Economy and Competitiveness (MINECO), 
and project 3E 310386 of the Instituto de Astrof\'isica de Canarias.
This research made use of APLpy, an open-source plotting package for Python hosted at http://aplpy.github.com.  
The scientific results 
reported in this article are based in part on data obtained from the Chandra Data Archive. 
This research has made use of software packages CIAO provided by the Chandra X-ray Centre (CXC) and FTOOLS provided 
by NASAS's HEASARC. We thank K. Iwasawa for helpful guidance on the X-Ray spectra. The authors thank the referee, 
F. Bournaud, for his helpful comments which have enabled us to make
useful improvements to the article.

     \bibliographystyle{mn2e}

\appendix
\section{HII regions properties}
\begin{table*}
\centering
\begin{tabular}{|c|c|c|c|c|c|c|c|c|c|}
\hline
N&RA&Dec &R &$log(L_{H\alpha})$&$n_e$&$\sigma_{v-nt}$&$log(M_{HII})$& $log(M_{gas})$ &$log(M_{vir})$ \\
 &     (hh:mm:ss)         &     ($^{\circ}$ $\mathrm{\prime}$ $\mathrm{\prime\prime}$)      &   (pc)   &     $\, \mathrm(erg/s)$                              &     $\, \mathrm(e^{-}/cm^3)    $                    &    $\, \mathrm{km/s}$                        & $  \, \mathrm{M_{\odot}}$ & $  \, \mathrm{M_{\odot}}$           &  $ \, \mathrm{M_{\odot}}$                                  \\
\hline
      1&          10          49     55.31&          32          59     26.25&         349&     40.45&      4.81&     34.93&      7.33&      8.11&      8.47\\
\hline
      2&          10          49     49.75&          32          59      3.28&         346&     40.33&      4.23&     23.73&      7.27&      8.04&      8.13\\
\hline
      3&          10          49     55.79&          32          59     27.78&         367&     40.35&      3.98&     24.82&      7.31&      8.09&      8.20\\
\hline
      4&          10          49     56.04&          32          59     28.19&         341&     40.15&      3.51&     25.28&      7.16&      7.94&      8.18\\
\hline
      5&          10          49     51.20&          32          59     11.56&         369&     39.95&      2.48&     18.80&      7.11&      7.89&      7.96\\
\hline
      6&          10          49     50.46&          32          59      2.51&         275&     39.78&      3.20&     12.29&      6.84&      7.62&      7.46\\
\hline
      7&          10          49     53.15&          32          59     10.47&         194&     39.26&      2.94&      8.49&      6.35&      7.13&      6.99\\
\hline
      8&          10          49     49.84&          32          58     53.78&         230&     39.48&      2.93&     14.92&      6.57&      7.35&      7.55\\
\hline
      9&          10          49     53.68&          32          59     10.77&         216&     39.19&      2.31&      8.23&      6.39&      7.16&      7.01\\
\hline
     10&          10          49     49.06&          32          59      4.43&         233&     39.33&      2.42&     10.70&      6.51&      7.29&      7.27\\
\hline
     11&          10          49     49.47&          32          58     46.37&         231&     39.40&      2.66&     12.71&      6.54&      7.32&      7.42\\
\hline
     12&          10          49     50.53&          32          59      0.92&         222&     39.36&      2.71&     15.56&      6.49&      7.27&      7.57\\
\hline
     13&          10          49     51.41&          32          59     12.13&         250&     39.44&      2.47&     12.07&      6.61&      7.39&      7.41\\
\hline
     14&          10          49     49.00&          32          58     59.17&         268&     39.43&      2.21&     13.47&      6.65&      7.43&      7.53\\
\hline
     15&          10          49     49.11&          32          59      2.93&         248&     39.42&      2.44&     11.83&      6.59&      7.37&      7.38\\
\hline
     16&          10          49     49.56&          32          59      0.12&         276&     39.52&      2.35&     18.17&      6.72&      7.49&      7.80\\
\hline
     17&          10          49     54.64&          32          59     26.43&         210&     39.33&      2.83&     18.58&      6.44&      7.22&      7.70\\
\hline
     18&          10          49     54.58&          32          59     25.76&         253&     39.44&      2.43&     17.26&      6.61&      7.39&      7.72\\
\hline
     19&          10          49     49.62&          32          58     56.36&         211&     39.26&      2.59&     14.08&      6.41&      7.19&      7.47\\
\hline
     20&          10          49     49.51&          32          58     50.54&         175&     39.13&      2.98&     13.07&      6.22&      7.00&      7.32\\
\hline
     21&          10          49     50.14&          32          58     59.53&         301&     39.69&      2.50&     18.18&      6.85&      7.63&      7.84\\
\hline
     22&          10          49     49.81&          32          59      7.64&         244&     39.48&      2.69&     15.31&      6.61&      7.39&      7.60\\
\hline
     23&          10          49     50.27&          32          58     49.88&         224&     39.26&      2.37&     10.52&      6.45&      7.22&      7.24\\
\hline
     24&          10          49     51.59&          32          59     12.95&         194&     39.01&      2.21&      8.39&      6.23&      7.01&      6.98\\
\hline
     25&          10          49     50.28&          32          59      7.45&         217&     39.20&      2.32&     12.78&      6.40&      7.18&      7.39\\
\hline
     26&          10          49     50.13&          32          58     55.11&         215&     39.25&      2.51&     10.73&      6.42&      7.20&      7.24\\
\hline
     27&          10          49     53.21&          32          59     12.76&         188&     38.99&      2.27&      8.46&      6.20&      6.98&      6.97\\
\hline
     28&          10          49     49.55&          32          58     52.00&         205&     39.24&      2.63&     13.05&      6.38&      7.16&      7.39\\
\hline
     29&          10          49     48.11&          32          58     25.08&         215&     39.07&      2.03&      9.35&      6.33&      7.10&      7.12\\
\hline
     30&          10          49     49.86&          32          58     36.86&         174&     39.06&      2.74&      9.86&      6.18&      6.96&      7.07\\
\hline
     31&          10          49     49.33&          32          58     49.27&         192&     39.07&      2.41&     10.12&      6.25&      7.03&      7.14\\
\hline
     32&          10          49     55.00&          32          59     26.62&         224&     39.31&      2.50&     23.64&      6.47&      7.25&      7.94\\
\hline
     33&          10          49     55.09&          32          59     24.51&         224&     39.32&      2.53&     23.43&      6.48&      7.25&      7.93\\
\hline
     34&          10          49     50.28&          32          59     11.59&         215&     39.18&      2.31&     13.66&      6.38&      7.16&      7.45\\
\hline
     35&          10          49     52.90&          32          59     11.06&         218&     39.10&      2.05&     12.28&      6.35&      7.13&      7.36\\
\hline
     36&          10          49     49.71&          32          58     49.21&         204&     39.12&      2.31&     11.56&      6.31&      7.09&      7.28\\
\hline
     37&          10          49     48.24&          32          58     41.64&         240&     39.08&      1.73&      9.09&      6.40&      7.18&      7.14\\
\hline

\end{tabular}
\caption{Physical properties derived as described in section $\S4$.}
\label{table:prop1}
\end{table*}  

\begin{table*}
\centering
\begin{tabular}{|c|c|c|c|c|c|c|c|c|c|}
\hline
N&RA&Dec &R &$log(L_{H\alpha})$&$n_e$&$\sigma_{v-nt}$&$log(M_{HII})$& $log(M_{gas})$ &$log(M_{vir})$ \\
 &     (hh:mm:ss)         &     ($^{\circ}$ $\mathrm{\prime}$ $\mathrm{\prime\prime}$)      &   (pc)   &     $\, \mathrm(erg/s)$                              &     $\, \mathrm(e^{-}/cm^3)    $                    &    $\, \mathrm{km/s}$                        & $  \, \mathrm{M_{\odot}}$ & $  \, \mathrm{M_{\odot}}$           &  $ \, \mathrm{M_{\odot}}$                                  \\

\hline
     38&          10          49     49.89&          32          58     56.42&         233&     39.27&      2.27&     19.17&      6.48&      7.26&      7.78\\
\hline
     39&          10          49     49.82&          32          59      9.27&         193&     38.97&      2.13&      8.31&      6.21&      6.98&      6.97\\
\hline
     40&          10          49     54.91&          32          59     25.59&         188&     39.15&      2.74&     15.06&      6.28&      7.06&      7.47\\
\hline
     41&          10          49     55.46&          32          59     23.19&         305&     39.48&      1.93&     21.98&      6.76&      7.54&      8.01\\
\hline
     42&          10          49     50.46&          32          58     55.12&         188&     39.02&      2.34&     12.68&      6.21&      6.99&      7.32\\
\hline
     43&          10          49     49.54&          32          58     57.67&         169&     38.90&      2.39&     12.76&      6.08&      6.86&      7.28\\
\hline
     44&          10          49     50.48&          32          58     57.56&         220&     39.09&      2.01&     14.24&      6.35&      7.13&      7.49\\
\hline
     45&          10          49     48.70&          32          58     40.99&         178&     38.77&      1.90&      9.17&      6.05&      6.83&      7.02\\
\hline
     46&          10          49     50.41&          32          58     53.38&         172&     38.90&      2.31&     10.25&      6.09&      6.87&      7.10\\
\hline
     47&          10          49     49.78&          32          58     38.61&         170&     38.89&      2.35&      9.75&      6.09&      6.86&      7.05\\
\hline
     48&          10          49     54.79&          32          59     22.64&         206&     39.07&      2.18&     16.10&      6.30&      7.08&      7.57\\
\hline
     49&          10          49     50.39&          32          59      9.86&         175&     38.98&      2.51&     13.58&      6.15&      6.93&      7.35\\
\hline
     50&          10          49     50.20&          32          58     57.23&         194&     39.07&      2.35&     15.78&      6.26&      7.04&      7.53\\
\hline
     51&          10          49     49.20&          32          58     39.37&         157&     38.70&      2.13&      9.54&      5.93&      6.71&      7.00\\
\hline
     52&          10          49     50.40&          32          58     51.48&         190&     38.93&      2.08&     12.25&      6.18&      6.95&      7.30\\
\hline
     53&          10          49     49.99&          32          59     11.89&         195&     38.87&      1.88&     15.03&      6.16&      6.94&      7.49\\
\hline
     54&          10          49     48.62&          32          58     45.00&         170&     38.72&      1.92&     10.71&      6.00&      6.77&      7.13\\
\hline
     55&          10          49     48.30&          32          58     44.37&         208&     38.94&      1.82&     11.50&      6.24&      7.02&      7.28\\
\hline
     56&          10          49     50.67&          32          59      7.39&         186&     38.83&      1.90&     10.27&      6.11&      6.89&      7.14\\
\hline
     57&          10          49     50.05&          32          59      7.23&         257&     39.24&      1.88&     18.10&      6.53&      7.31&      7.77\\
\hline
     58&          10          49     49.37&          32          58     55.33&         198&     38.86&      1.80&     10.18&      6.17&      6.94&      7.16\\
\hline
     59&          10          49     49.23&          32          58     46.51&         184&     38.81&      1.90&     10.05&      6.09&      6.87&      7.11\\
\hline
     60&          10          49     46.70&          32          58     24.66&         124&     38.43&      2.21&      8.13&      5.65&      6.43&      6.76\\
\hline
     61&          10          49     48.53&          32          58     42.07&         204&     38.90&      1.81&      8.40&      6.21&      6.99&      7.00\\
\hline
     62&          10          49     49.94&          32          58     46.90&         157&     38.63&      1.97&     11.16&      5.90&      6.68&      7.14\\
\hline
     63&          10          49     50.65&          32          59      8.92&         179&     38.77&      1.89&     11.05&      6.06&      6.83&      7.18\\
\hline
     64&          10          49     49.89&          32          58     48.14&         145&     38.61&      2.15&     12.11&      5.84&      6.62&      7.17\\
\hline
     65&          10          49     50.89&          32          59      2.74&         120&     38.40&      2.27&      8.90&      5.61&      6.39&      6.82\\
\hline
     66&          10          49     49.09&          32          58     40.04&         163&     38.70&      2.02&     10.91&      5.96&      6.74&      7.13\\
\hline
     67&          10          49     50.57&          32          59     11.30&         162&     38.73&      2.10&     15.49&      5.97&      6.75&      7.43\\
\hline
     68&          10          49     49.04&          32          58     38.75&         124&     38.43&      2.21&     10.56&      5.65&      6.43&      6.99\\
\hline
     69&          10          49     49.64&          32          58     36.46&         122&     38.45&      2.32&     10.38&      5.65&      6.43&      6.96\\
\hline
     70&          10          49     49.65&          32          58     37.91&         120&     38.43&      2.33&     11.53&      5.62&      6.40&      7.04\\
\hline
     71&          10          49     49.83&          32          58     41.70&         134&     38.47&      2.07&      9.71&      5.71&      6.49&      6.94\\
\hline
     72&          10          49     50.44&          32          59      6.73&         197&     38.82&      1.74&      8.32&      6.14&      6.92&      6.98\\
\hline
     73&          10          49     49.74&          32          58     34.63&         171&     38.59&      1.65&      8.65&      5.94&      6.72&      6.95\\
\hline
     74&          10          49     50.08&          32          58     48.78&         145&     38.66&      2.28&     16.14&      5.86&      6.64&      7.42\\
\hline

\end{tabular}
\caption{Physical properties derived as described in section $\S4$.}
\label{table:prop2}
\end{table*}  
\begin{table*}
\centering
\begin{tabular}{|c|c|c|c|c|c|c|c|c|c|}
\hline
N&RA&Dec &R &$log(L_{H\alpha})$&$n_e$&$\sigma_{v-nt}$&$log(M_{HII})$& $log(M_{gas})$ &$log(M_{vir})$ \\
 &     (hh:mm:ss)         &     ($^{\circ}$ $\mathrm{\prime}$ $\mathrm{\prime\prime}$)      &   (pc)   &     $\, \mathrm(erg/s)$                              &     $\, \mathrm(e^{-}/cm^3)    $                    &    $\, \mathrm{km/s}$                        & $  \, \mathrm{M_{\odot}}$ & $  \, \mathrm{M_{\odot}}$           &  $ \, \mathrm{M_{\odot}}$                                  \\

\hline
     75&          10          49     50.16&          32          58     46.12&         157&     38.63&      1.96&     11.86&      5.90&      6.68&      7.19\\
\hline
     76&          10          49     48.64&          32          58     35.73&         153&     38.49&      1.74&      8.90&      5.81&      6.59&      6.93\\
\hline
     77&          10          49     54.71&          32          59     20.25&         177&     38.61&      1.60&      9.66&      5.97&      6.75&      7.06\\
\hline
     78&          10          49     56.46&          32          59     27.88&         209&     38.64&      1.28&     10.73&      6.09&      6.87&      7.23\\
\hline
     79&          10          49     54.47&          32          59     22.46&         235&     38.91&      1.48&     11.13&      6.31&      7.08&      7.31\\
\hline
     80&          10          49     48.97&          32          58     39.43&         175&     38.65&      1.71&     11.17&      5.98&      6.76&      7.18\\
\hline
     81&          10          49     49.94&          32          58     50.32&         181&     38.71&      1.73&     11.40&      6.03&      6.81&      7.21\\
\hline
     82&          10          49     49.97&          32          58     44.58&         164&     38.52&      1.62&      8.11&      5.88&      6.65&      6.88\\
\hline
     83&          10          49     50.26&          32          58     46.81&         169&     38.62&      1.73&     12.52&      5.94&      6.72&      7.27\\
\hline
     84&          10          49     50.70&          32          59     11.56&         252&     38.91&      1.33&     18.57&      6.35&      7.13&      7.78\\
\hline
     85&          10          49     49.96&          32          58     40.79&         137&     38.38&      1.81&      9.75&      5.69&      6.47&      6.96\\
\hline
     86&          10          49     49.33&          32          59      6.12&         237&     38.83&      1.32&     15.95&      6.27&      7.05&      7.62\\
\hline
     87&          10          49     55.77&          32          59     23.54&         232&     38.90&      1.48&     22.63&      6.29&      7.07&      7.92\\
\hline
     88&          10          49     50.73&          32          59      3.46&         175&     38.59&      1.60&     10.46&      5.95&      6.73&      7.13\\
\hline
     89&          10          49     49.49&          32          59      7.02&         230&     38.79&      1.33&     18.43&      6.23&      7.01&      7.74\\
\hline
     90&          10          49     49.34&          32          58     59.00&         183&     38.57&      1.46&     15.16&      5.97&      6.75&      7.47\\
\hline
     91&          10          49     49.10&          32          58     37.52&         152&     38.47&      1.70&     14.98&      5.80&      6.58&      7.38\\
\hline
     92&          10          49     55.02&          32          59     29.79&         206&     38.60&      1.25&     22.58&      6.06&      6.84&      7.87\\
\hline
     93&          10          49     49.17&          32          58     35.93&         117&     38.11&      1.67&      8.98&      5.45&      6.23&      6.82\\
\hline
     94&          10          49     51.09&          32          59      8.08&         132&     38.20&      1.55&     11.23&      5.57&      6.35&      7.06\\
\hline
     95&          10          49     50.82&          32          59      0.18&         108&     38.00&      1.66&      8.66&      5.34&      6.12&      6.75\\
\hline
     96&          10          49     50.68&          32          59     14.61&          97&     37.88&      1.69&      9.10&      5.22&      5.99&      6.75\\
\hline
     97&          10          49     49.74&          32          58     43.64&         120&     38.12&      1.64&      9.22&      5.47&      6.25&      6.85\\
\hline
     98&          10          49     49.90&          32          59     12.82&         165&     38.40&      1.40&     12.20&      5.82&      6.60&      7.23\\
\hline
     99&          10          49     50.02&          32          58     50.19&         126&     38.14&      1.56&      8.28&      5.51&      6.29&      6.78\\
\hline
    100&          10          49     49.79&          32          59     11.59&         117&     38.06&      1.60&     14.71&      5.42&      6.20&      7.25\\
\hline
    101&          10          49     49.26&          32          58     58.69&          69&     37.72&      2.33&     19.02&      4.91&      5.69&      7.24\\
\hline
    102&          10          49     48.99&          32          58     45.30&          91&     37.87&      1.85&      9.74&      5.17&      5.95&      6.78\\
\hline
    103&          10          49     48.78&          32          58     43.11&         140&     38.22&      1.45&     12.54&      5.62&      6.40&      7.19\\
\hline
    104&          10          49     54.70&          32          59     31.24&          62&     37.47&      2.10&     18.43&      4.71&      5.49&      7.16\\
\hline
    105&          10          49     55.53&          32          59     32.69&         171&     38.38&      1.29&     20.44&      5.83&      6.61&      7.70\\
\hline
    106&          10          49     48.20&          32          58     51.53&         167&     38.36&      1.31&     11.80&      5.81&      6.59&      7.21\\
\hline
    107&          10          49     54.96&          32          59     20.84&         142&     38.20&      1.39&     10.84&      5.62&      6.40&      7.07\\
\hline
    108&          10          49     50.71&          32          58     54.80&          98&     37.92&      1.74&     13.00&      5.24&      6.02&      7.06\\
\hline

\end{tabular}
\caption{Physical properties derived as described in section $\S4$.}
\label{table:prop3}
\end{table*}  

\label{lasttpage}
\end{document}